\pdfoutput=1
\documentclass[11pt,letterpaper]{article}

\usepackage[T1]{fontenc}
\usepackage{mathpazo}
\usepackage[margin=1in]{geometry}
\usepackage{amsmath,amssymb}
\usepackage{graphicx}
\usepackage{booktabs}
\usepackage{float}
\usepackage{changepage}
\usepackage[font=small,labelfont=bf,labelsep=period]{caption}
\usepackage[numbers,sort&compress]{natbib}
\usepackage{tikz}
\usepackage{microtype}
\PassOptionsToPackage{hyphens}{url}
\usepackage[colorlinks=true,allcolors=blue!55!black]{hyperref}
\makeatletter
\newlength{\extralength}
\newcommand{\Title}[1]{\gdef\sa@Title{#1}}
\newcommand{\Author}[1]{\gdef\sa@Author{#1}}
\newcommand{\AuthorNames}[1]{\gdef\sa@AuthorNames{#1}}
\newcommand{\address}[1]{\gdef\sa@address{#1}}
\newcommand{\corres}[1]{\gdef\sa@corres{#1}}
\renewcommand{\abstract}[1]{\gdef\sa@abstract{#1}}
\let\endabstract\relax
\newcommand{\keyword}[1]{\gdef\sa@keyword{#1}}
\newcommand{\sa@frontmatter}{%
  \begingroup
  \begin{flushleft}
  {\LARGE\bfseries\boldmath \sa@Title\par}
  \vspace{12pt}
  {\large\bfseries\boldmath \sa@Author\par}
  \vspace{8pt}
  {\small \sa@address\\ {*} \quad \sa@corres\par}
  \vspace{6pt}
  {\small \@date\par}
  \end{flushleft}
  \vspace{2pt}
  \noindent{\small\textbf{Abstract:} \sa@abstract\par}
  \vspace{8pt}
  \noindent{\small\textbf{Keywords:} \sa@keyword\par}
  \vspace{12pt}
  \endgroup}
\AtBeginDocument{\sa@frontmatter}
\newcommand{\sa@statement}[2]{\par\vspace{6pt}\noindent{\small\textbf{#1:} #2\par}}
\newcommand{\authorcontributions}[1]{\sa@statement{Author Contributions}{#1}}
\newcommand{\funding}[1]{\sa@statement{Funding}{#1}}
\newcommand{\institutionalreview}[1]{\sa@statement{Institutional Review Board Statement}{#1}}
\newcommand{\informedconsent}[1]{\sa@statement{Informed Consent Statement}{#1}}
\newcommand{\dataavailability}[1]{\sa@statement{Data Availability Statement}{#1}}
\newcommand{\acknowledgments}[1]{\sa@statement{Acknowledgments}{#1}}
\newcommand{\conflictsofinterest}[1]{\sa@statement{Conflicts of Interest}{#1}}
\newcommand{\abbreviations}[2]{\par\vspace{12pt}\noindent{\large\bfseries #1\par}\vspace{3pt}\noindent{\small\ignorespaces #2\par}}
\newcommand{\reftitle}[1]{\renewcommand{\refname}{#1}}
\newcommand{\PublishersNote}{}
\renewcommand{\bibfont}{\small}
\makeatother

\usepackage{pgf}
\usetikzlibrary{decorations.pathmorphing,decorations.markings,arrows.meta}
\tikzset{
  fermion/.style={thick,postaction={decorate},decoration={markings,mark=at position 0.55 with {\arrow{Stealth[length=2mm]}}}},
  boson/.style={thick,decorate,decoration={snake,amplitude=1.3pt,segment length=5pt}},
  gluon/.style={thick,decorate,decoration={coil,aspect=0.7,segment length=3.2pt,amplitude=2pt}},
  vtx/.style={circle,fill,inner sep=0.9pt},
  lbl/.style={font=\small},
}
\providecommand{\mathdefault}[1]{#1}% used by the embedded PGF figures
\newcommand{\ttbar}{t\bar t}
\newcommand{\mtt}{m_{t\bar t}}
\newcommand{\etat}{\eta_t}
\date{November 24, 2026}

\Title{Top-Quark Physics after Evidence for Toponium}
\Author{Howard Baer $^{1}$ and Vernon Barger $^{2,}$*}
\AuthorNames{Howard Baer and Vernon Barger}
\address{%
$^{1}$ \quad Homer L. Dodge Department of Physics and Astronomy, University of Oklahoma, Norman, OK 73019, USA; baer@ou.edu\\
$^{2}$ \quad Department of Physics, University of Wisconsin--Madison, Madison, WI 53706, USA}
\corres{Correspondence: barger@pheno.wisc.edu}

\abstract{The top quark decays before it hadronizes, and toponium, the $t\bar t$ bound state, was long thought unobservable as a resonance. CMS and ATLAS each report an excess of top-quark pairs at threshold, above five standard deviations over fixed-order quantum chromodynamics (QCD), and CMS confirms it in a second decay channel. Its angular distributions match the color-singlet pseudoscalar quasi-bound state $\eta_t$ of nonrelativistic QCD, although an additional pseudoscalar boson is not excluded and higher-order fixed-order QCD accounts for part of the enhancement. We review the theory, ATLAS's nominal Green's-function signal model, and the measurements, whose cross sections range from about 3 to 13 pb. Part of the spread comes from the signal model and part persists between analyses using the same model. We discuss consequences for entanglement, the top mass, and heavy-pseudoscalar searches. We then review the top-quark program, from mass definitions to electroweak precision, flavor, the charge-conjugation and parity (CP) structure of the top Yukawa coupling, supersymmetry, multi-top production, and effective field theory. We close with prospects at the High-Luminosity LHC and future colliders, where an $e^+e^-$ threshold scan can determine the top-quark mass in a short-distance scheme to a few tens of MeV.}

\keyword{toponium; top quark; nonrelativistic QCD; Green's function; spin correlation; entanglement; top Yukawa coupling; CP violation; top-quark mass; LHC}

\begin{document}

\section{Introduction}\label{sec:intro}

\subsection{Historical background}\label{sec:history}

The top quark has a mass $m_t\simeq 172.5$ GeV and a width $\Gamma_t\simeq 1.33$ GeV. Its lifetime, $1/\Gamma_t\simeq5\times10^{-25}$ s, is shorter than the time scale $1/\Lambda_{\rm QCD}\simeq3\times10^{-24}$ s on which confinement acts, with $\Lambda_{\rm QCD}\approx0.2$ GeV the scale parameter of quantum chromodynamics (QCD)~\cite{Bigi:1986jk}. It therefore decays before it can hadronize and passes its spin to its decay products, and is the only quark that can be studied bare. The same fact led to the expectation~\cite{Bigi:1986jk} that toponium, the $\ttbar$ analog of charmonium and bottomonium, would never be seen as a resonance, at least at a hadron collider. Where the scheme is not specified, $m_t$ denotes the top-quark mass at the nominal value of 172.5 GeV used in the experimental analyses. Where it matters, we write $m_t^{\rm MC}$ for the mass parameter of the event generators determined in direct measurements, $m_t^{\rm pole}$ for the pole mass, and $m_t(\mu)$ for the $\overline{\rm MS}$ mass at the scale $\mu$ (Section~\ref{sec:props}).

Toponium was studied intensively in the 1980s. The top mass was then unknown, and narrow bound states of heavy quarks, toponium among them, looked like a promising source of Higgs signals~\cite{Barger:1986nn,Kuhn:1987ty}. Fadin and Khoze showed that, for a sufficiently heavy top quark, the width acts as an infrared cutoff that smears the would-be resonances into a broad enhancement at threshold~\cite{Fadin:1987wz,Strassler:1990nw}. Their analysis was extended to hadron colliders~\cite{Fadin:1990wx} and developed into a quantitative theory of threshold production~\cite{Sumino:1992ai,Strassler:1990nw}.

Hadron-collider searches for the top quark began at the CERN $p\bar p$ collider. Barger, Martin, and Phillips showed that the transverse mass of the electron and neutrino, which they had analyzed for $W\to e\nu$ decays~\cite{Barger:1983wf}, also identifies semileptonic decays of a heavy quark and bounds its mass~\cite{Barger:1983jx}. They also introduced the cluster transverse mass, which combines a cluster of visible particles with the missing transverse momentum~\cite{Barger:1983jx,Barger:1983qr,Barger:1984sm}. For $W\to t\bar b$, its distribution for the $t\bar b$ system has a Jacobian peak just below $M_W$~\cite{Barger:1984sm}. The variable was later applied to Higgs-boson decays to gauge-boson pairs~\cite{Barger:1987du}.

In 1984 UA1 reported lepton-plus-jets events consistent with $W\to t\bar b$ and $30<m_t<50$ GeV~\cite{UA1:1984nbd}. Larger UA1 samples found no signal and set $m_t>60$ GeV~\cite{UA1:1990rck}.

For the Tevatron, attention turned to heavier top quarks decaying to real $W$ bosons. Baer, Barger, and Phillips proposed selection cuts for these heavier top quarks, in signatures with a single isolated lepton plus high jet multiplicity and with two isolated leptons plus jets~\cite{Baer:1988kq}. The variable $H_T=\sum E_T$, the scalar sum of the transverse energies of all jets, was suggested to separate the $\ttbar$ signal from the $W$+multijet background, and the top mass could be reconstructed by combining $b$ jets with the jets from $W\to q\bar q'$. A Tevatron reach of $m_t\approx180$ GeV was found for 100 pb$^{-1}$ of data. In 1995 the CDF collaboration was able to make use of $b$-jet micro-vertex tagging to aid in the top discovery~\cite{CDF:1995wbb}. The D0 analysis lacked $b$-vertex tagging but was nonetheless able to isolate a top-quark signal by judicious use of the $H_T$ variable~\cite{D0:1995jca}.

\subsection{The threshold observations}\label{sec:observations}

Detailed predictions for the LHC followed~\cite{Hagiwara:2008df,Kiyo:2008bv,Sumino:2010bv}, and the LHC Run~2 data provided a sample large enough to test them. CMS reported a significant excess of $\ttbar$ events at threshold in the dilepton channel, consistent with a color-singlet pseudoscalar quasi-bound state~\cite{CMS:2025kzt}. ATLAS confirmed the excess~\cite{ATLAS:2025conf,ATLAS:2026toponium}, and CMS has since observed it in the lepton+jets channel~\cite{CMS:2026lj}. These results are the focus of this article.

The guest editors of this Special Issue have contributed to this development and to related work on the top Yukawa coupling. The nominal signal model in the ATLAS observation and the simplified pseudoscalar model used by CMS both originate in work of Fuks, Hagiwara, Ma, and Zheng~\cite{Fuks:2021xje,Fuks:2024yjj}. Zheng, with Hagiwara and Barger, has also worked extensively on the charge-conjugation and parity (CP) structure of the top-quark Yukawa coupling~\cite{Barger:2018tqn,Barger:2019ccj,Barger:2023wbg}, and Lewis, Zheng, and collaborators have studied its measurement at a muon collider~\cite{Cassidy:2023lwd}. We highlight these contributions where they arise.

\subsection{Scope and outline}\label{sec:outline}

Section~\ref{sec:threshold} summarizes the physics of the $\ttbar$ threshold in nonrelativistic QCD (NRQCD). Section~\ref{sec:exp} reviews the experimental observations. Section~\ref{sec:model} describes how threshold effects are simulated and why the extracted cross sections depend on the model. Section~\ref{sec:implications} discusses the implications for QCD, spin correlations, the top mass and Yukawa coupling, and new physics.

Section~\ref{sec:props} reviews the top-quark mass and width. Section~\ref{sec:prodcoup} covers single-top production, the electroweak couplings, the charge asymmetry, and top quarks at high momentum, in the forward region, and in nuclear collisions. Section~\ref{sec:ewsb} explains why the top mass plays a large role in electroweak symmetry breaking and in rare processes. Section~\ref{sec:yukawa} turns to the CP structure of the top Yukawa coupling. Section~\ref{sec:bsm} covers the third row of the Cabibbo--Kobayashi--Maskawa (CKM) matrix, including the determination of the CP-violating phase from $|V_{td}|$, and global effective-field-theory fits. Section~\ref{sec:susyhiggs} discusses the role of the top quark in the supersymmetric Higgs mass and in radiative natural supersymmetry. Section~\ref{sec:multitop} discusses three- and four-top production. Section~\ref{sec:future} outlines future directions, and Section~\ref{sec:concl} concludes.

Topics beyond the scope of this article include vector-like top partners, which composite-Higgs and little-Higgs solutions of the hierarchy problem predict~\cite{Panico:2015jxa,Bellazzini:2014yua,Schmaltz:2005ky}. Their mixing with the top quark, decays, and production are reviewed in Refs.~\cite{Aguilar-Saavedra:2013qpa,Alves:2023ufm}, and the LHC searches in Refs.~\cite{CMS:2024vlqreview,ATLAS:2024vlqsingle,CMS:2022fck,Benbrik:2024fku}. Lewis and collaborators have contributed to this subject by studying the effect of top partners on Higgs production~\cite{Chen:2014xwa}. They have also studied several less conventional signatures, namely loop-induced single production with a top quark through a singlet scalar~\cite{Kim:2018mks}, pair production with decays to $tg$ and $t\gamma$~\cite{Alhazmi:2018whk}, and decays to dark photons~\cite{Kim:2019oyh}.

\section{The Top-Quark Threshold in Nonrelativistic QCD}\label{sec:threshold}

\subsection{Scales}\label{sec:scales}

Near threshold the $\ttbar$ system is nonrelativistic. For a pair of invariant mass $M=\mtt$, the top-quark velocity in the pair rest frame is

\begin{equation}
v=\sqrt{1-\frac{4m_t^2}{M^2}},\qquad E\equiv M-2m_t\simeq m_t v^2 ,
\label{eq:v}
\end{equation}

so that $v\simeq 0.17$ at $M=350$ GeV. NRQCD and potential NRQCD organize the dynamics according to the hierarchy of three scales, the hard scale $m_t$, the soft (momentum) scale $m_t v$, and the ultrasoft (energy) scale $m_t v^2$~\cite{Caswell:1985ui,Bodwin:1994jh,Brambilla:2004jw}. A slowly moving top and antitop exchange potential gluons, generating terms of order $(\alpha_s/v)^n$ in the cross section, where $\alpha_s$ is the strong coupling. These terms are not small near threshold and must be resummed.

At leading order the potential at top--antitop separation $r$ is Coulombic and depends on the color state of the pair,

\begin{equation}
V^{[1]}(r)=-\frac{C_F\alpha_s}{r},\qquad V^{[8]}(r)=+\frac{(C_A/2-C_F)\,\alpha_s}{r}=+\frac{\alpha_s}{6r},
\label{eq:potential}
\end{equation}

with $C_F=4/3$ and $C_A=3$. The color-singlet channel is attractive and the color-octet channel repulsive. In the singlet channel the leading-order spectrum is hydrogen-like, with principal quantum number $n$,

\begin{equation}
E_n=-\frac{(C_F\alpha_s)^2\,m_t}{4n^2},\qquad r_B=\frac{2}{C_F\alpha_s m_t}.
\label{eq:hydrogen}
\end{equation}

For $\alpha_s$ between 0.13 and 0.15, evaluated near the soft scale, $|E_1|\simeq 1.3$--$1.7$ GeV and $r_B\simeq 0.012$ fm. These are rough, leading-order estimates. The higher-order potential and the running of $\alpha_s$ shift them, and the experimental signal models place the ground state about 2 GeV below $2m_t$~\cite{ATLAS:2026toponium}. Such estimates also depend on how the QCD scale and the quark mass are fixed. Potential models for charmonium and bottomonium add linear confinement to the short-distance Coulomb term, as in the Cornell potential~\cite{Eichten:1974af,Eichten:1975ag,Eichten:1978tg,Eichten:1979ms}. The Wisconsin potential connects the two regimes. It contains the full content of perturbative QCD at short distances and linear confinement at large distances, with the intermediate region determined from the observed levels~\cite{Hagiwara:1983qw,Miller:1981mj}, and it has also been used with a relativistic kinetic energy~\cite{Jacobs:1986gv}. Fits of this potential to the charmonium and bottomonium spectra show that a larger QCD scale parameter, combined with heavier quark masses, leaves the observed levels and leptonic widths nearly unchanged~\cite{Hagiwara:1983qw}.

\begin{figure}[H]
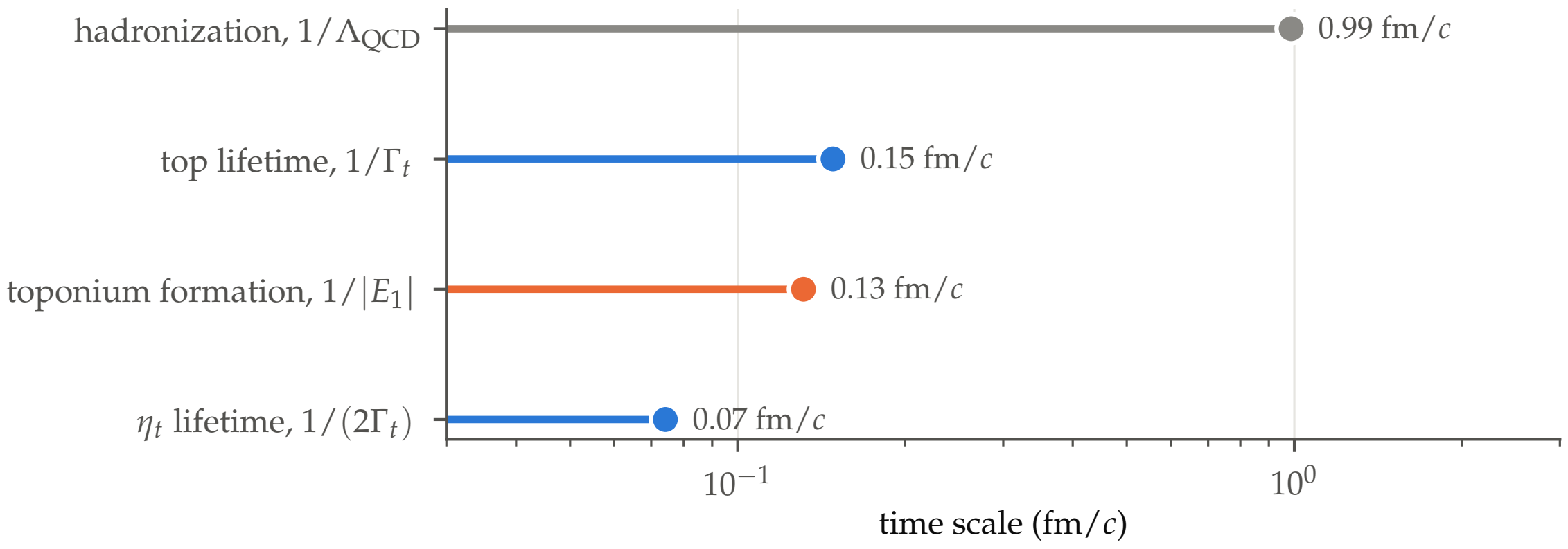

\begin{adjustwidth}{-\extralength}{0cm}
\centering
% ==== embedded figure fig_scales: generated by make_figures.py (matplotlib PGF backend); regenerate rather than edit ====
\begingroup%
\makeatletter%
%
\makeatother%
\endgroup%
% ==== end of embedded figure fig_scales ====
\caption{Time scales of the $\ttbar$ threshold region, obtained from $\Gamma_t=1.33$ GeV, the Coulomb binding energy of Equation~(\ref{eq:hydrogen}) with $\alpha_s=0.14$, and $\Lambda_{\rm QCD}\approx0.2$ GeV. Toponium forms and decays on the same time scale, and both occur well before hadronization.}
\label{fig:scales}
\end{adjustwidth}
\end{figure}

The formation time $1/|E_1|\simeq 0.13$ fm/$c$ is comparable to the lifetime $1/\Gamma_t\simeq 0.15$ fm/$c$ (Figure~\ref{fig:scales}). Its levels have widths ($\Gamma_{\etat}\simeq 2\Gamma_t$ for the pseudoscalar ground state $\etat$) larger than their spacing, so the $\ttbar$ threshold at the LHC shows a single broad enhancement just below $2m_t$ rather than a series of narrow peaks. Current detectors cannot resolve structure within this enhancement, since the $\mtt$ resolution near threshold is about 22\% in the ATLAS dilepton analysis~\cite{ATLAS:2026toponium}. Because its levels overlap, the system is often called a quasi-bound state~\cite{CMS:2025kzt,ATLAS:2026toponium}, and the name ``toponium'' for the observed effect should be read in that sense~\cite{Llanes-Estrada:2024phk}.

Top-flavored hadrons, by contrast, do not form. A $t\bar q$ meson, bound by confinement, would have level spacings of a few hundred MeV, set by the light quark and independent of the heavy-quark mass, and would need a time of order $1/\Lambda_{\rm QCD}\approx1$ fm/$c$ to form. Its levels would be several times broader than their spacing~\cite{Bigi:1986jk}. Toponium is the exception because its binding is a short-distance Coulomb effect whose strength grows with $m_t$.

\subsection{The Green's function}\label{sec:green}

The $(\alpha_s/v)^n$ terms are resummed by the nonrelativistic Green's function, which satisfies~\cite{Kiyo:2008bv,Sumino:2010bv}

\begin{equation}
\left[-\frac{\nabla^2}{m_t}+V^{[c]}(\vec r)-\left(E+i\Gamma_t\right)\right]G^{[c]}(\vec r;E+i\Gamma_t)=\delta^{(3)}(\vec r),\qquad c=1,8 .
\label{eq:green}
\end{equation}

The width enters as an imaginary part of the energy. Because $\Gamma_t\gg\Lambda_{\rm QCD}$, this makes the problem perturbative. For S-wave production the partonic cross section near threshold factorizes into a hard coefficient times ${\rm Im}\,G^{[c]}(0;E+i\Gamma_t)$. The momentum-space Green's function $\tilde G^{[c]}(\vec p;E+i\Gamma_t)$ gives the distribution of the top-quark momentum $\vec p$ in the pair rest frame~\cite{Sumino:1992ai,Sumino:2010bv}.

\begin{figure}[H]
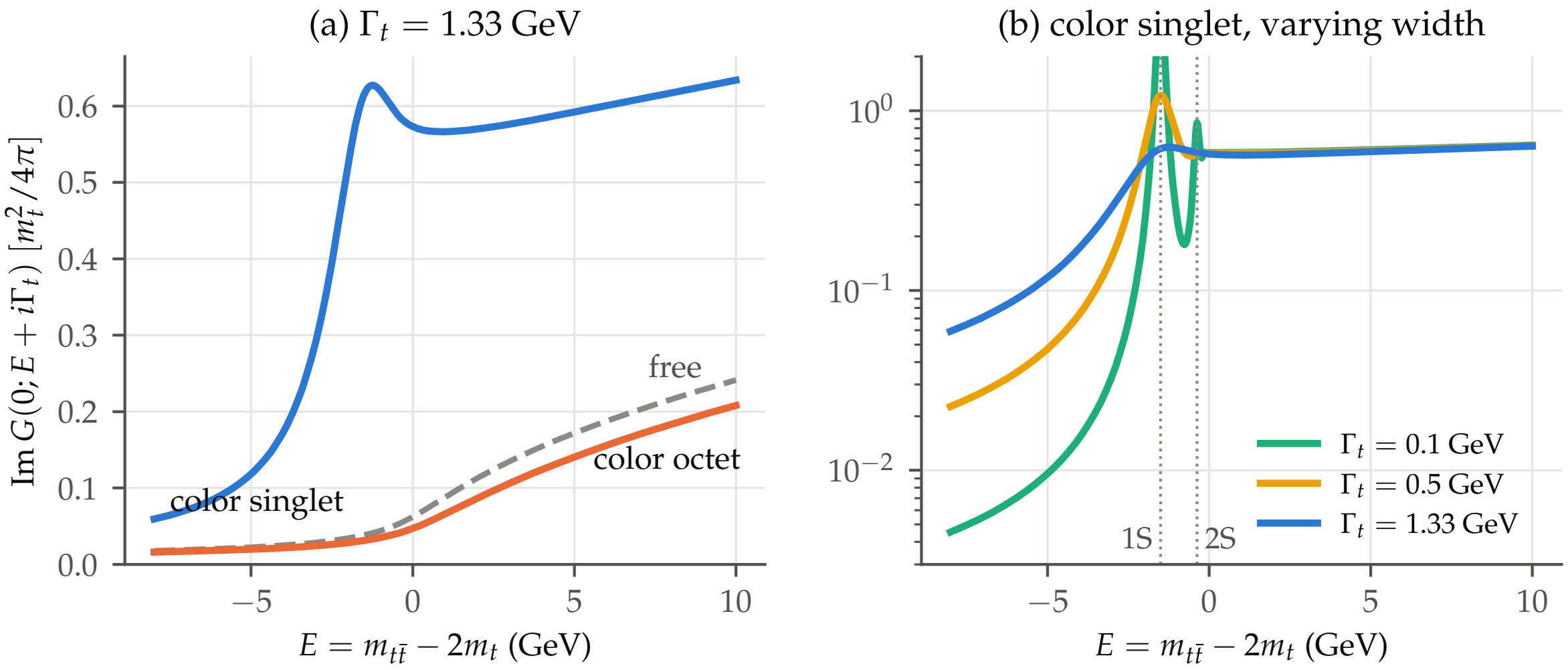

\begin{adjustwidth}{-\extralength}{0cm}
\centering
% ==== embedded figure fig_green: generated by make_figures.py (matplotlib PGF backend); regenerate rather than edit ====
\begingroup%
\makeatletter%
%
\makeatother%
\endgroup%
% ==== end of embedded figure fig_green ====
\caption{Imaginary part of the leading-order Coulomb Green's function at the origin, Equation~(\ref{eq:green}), proportional to the S-wave threshold cross section, as a function of $E=\mtt-2m_t$. (\textbf{a}) Color-singlet, free, and color-octet channels for $\Gamma_t=1.33$ GeV. (\textbf{b}) The color-singlet function for three values of the width. The dotted lines mark the Coulomb 1S and 2S energies of Equation~(\ref{eq:hydrogen}), with $E_1=-1.5$ GeV. For the physical width the levels merge into a single enhancement below threshold. The curves use $m_t=172.5$ GeV and a fixed $\alpha_s=0.14$, without higher-order potential corrections, the running of $\alpha_s$, or the smearing by parton luminosities and detector resolution. They illustrate the mechanism rather than a prediction.}
\label{fig:green}
\end{adjustwidth}
\end{figure}

Figure~\ref{fig:green} shows how the potential shapes the cross section. Without a potential, ${\rm Im}\,G$ reduces to the phase-space factor proportional to $v$. The attractive singlet potential produces smeared bound-state structure below $E=0$ and a Sommerfeld enhancement above it~\cite{Sommerfeld:1931qaf,Sakharov:1948plh}. The repulsive octet potential suppresses the rate. Predictions for LHC distributions at next-to-leading order in the potential, combined with threshold resummation of soft and collinear radiation, are available~\cite{Kiyo:2008bv,Garzelli:2024uhe,Garzelli:2026thr}, and the Coulomb resummation has also been matched to fixed-order NNLO QCD~\cite{Ju:2020otc}.

\subsection{Production channels}\label{sec:channels}

\begin{figure}[H]
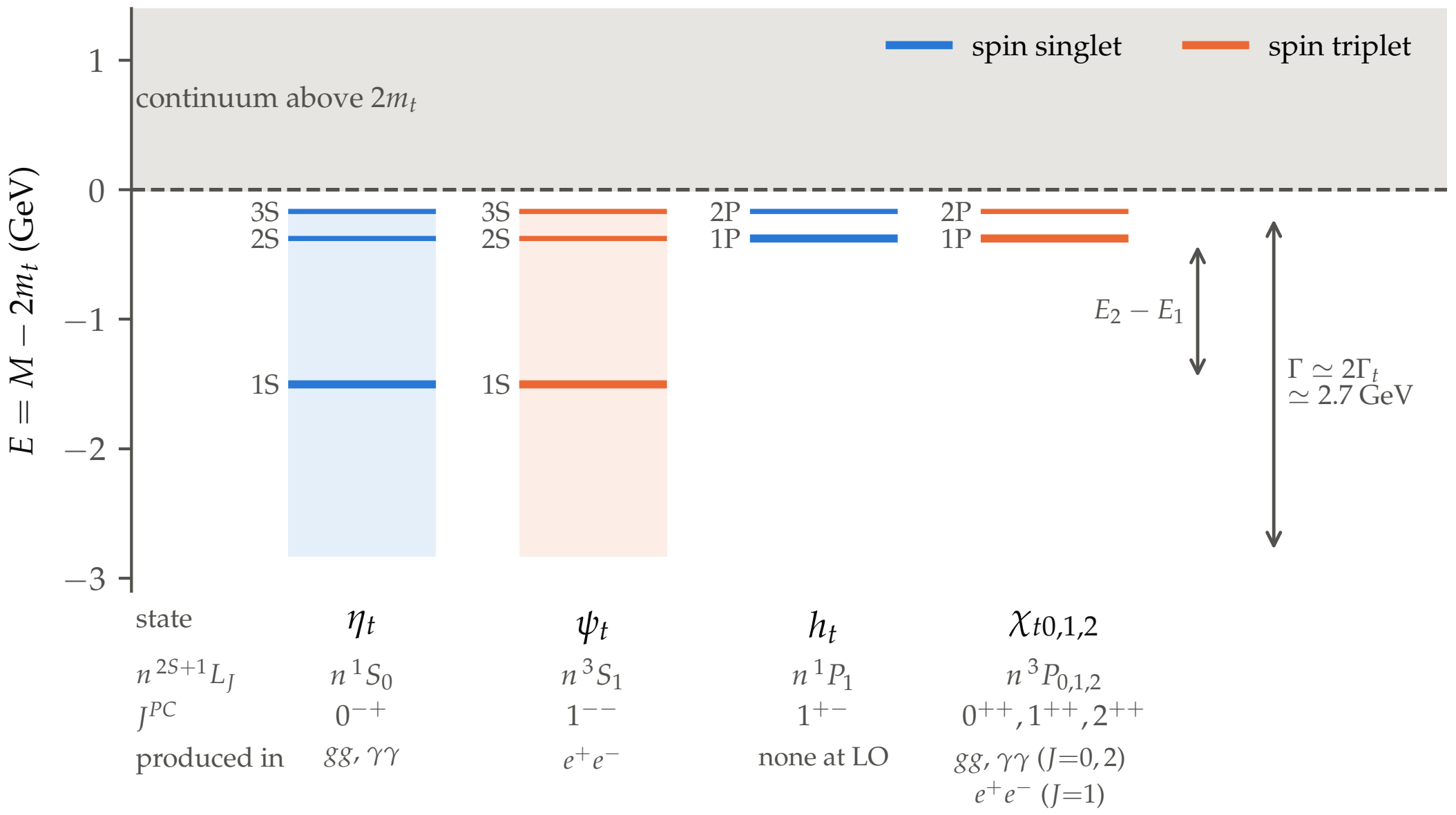

\begin{adjustwidth}{-\extralength}{0cm}
\centering
% ==== embedded figure fig_levels: generated by make_figures.py (matplotlib PGF backend); regenerate rather than edit ====
\begingroup%
\makeatletter%
%
\makeatother%
\endgroup%
% ==== end of embedded figure fig_levels ====
\caption{Color-singlet toponium levels and nomenclature in the Coulomb approximation of Equation~(\ref{eq:hydrogen}) with $\alpha_s=0.14$ and $m_t=172.5$ GeV, which gives $E_1\simeq-1.5$ GeV, $E_2\simeq-0.38$ GeV, and $E_3\simeq-0.17$ GeV. States are labeled by the radial quantum number $n$ and the orbital angular momentum, so that the 1P levels have principal quantum number 2, the index of $E_n$ in Equation~(\ref{eq:hydrogen}), and lie at $E_2$. Spin-singlet states are shown in blue and spin-triplet states in orange. Singlet and triplet levels are degenerate at this order, and higher-order corrections to the potential shift all levels. The shaded bands show the width $\Gamma\simeq2\Gamma_t\simeq2.7$ GeV of the 1S states. It exceeds the 1S--2S spacing, so the levels merge into a single enhancement below threshold (Figure~\ref{fig:green}). The bottom row lists the leading-order production channels (Table~\ref{tab:spectroscopy}). The ${}^1P_1$ state $h_t$ has charge-conjugation parity $C=-1$ and is not produced at leading order in $gg$, $\gamma\gamma$, or $e^+e^-$ collisions.}
\label{fig:levels}
\end{adjustwidth}
\end{figure}

At the LHC, $\ttbar$ production near threshold is dominated by gluon fusion. A color-singlet S-wave pair produced from two gluons must be in the ${}^1S_0^{[1]}$ configuration, i.e., the pseudoscalar $\etat$. In the notation ${}^{2S+1}L_J^{[c]}$, $S$, $L$, and $J$ are the total spin, orbital angular momentum, and total angular momentum of the pair, and the superscript $[c]=[1]$ or $[8]$ labels the color-singlet or color-octet state, as for the potentials of Equation~(\ref{eq:potential}). The vector state $\psi_t$ (${}^3S_1^{[1]}$) cannot be produced from two gluons at leading order, by the same selection rule that forbids $J/\psi\to gg$. In quark--antiquark annihilation the color-singlet channel is absent at leading order~\cite{Kiyo:2008bv,Fuks:2021xje,Bai:2025toponia}. The LHC signal is expected to be mainly pseudoscalar and spin singlet. P-wave and color-octet contributions are smaller but needed for a complete description~\cite{ATLAS:2026toponium,Garzelli:2026thr}. Figure~\ref{fig:levels} shows the lowest color-singlet levels, their nomenclature, and the channels in which they are produced.

The spin-singlet character has a direct observable consequence. Let $\varphi$ be the angle between the two charged-lepton directions, each measured in the rest frame of its parent top or antitop. The normalized distribution is

\begin{equation}
\frac{1}{\sigma}\frac{d\sigma}{d\cos\varphi}=\frac{1}{2}\left(1-D\cos\varphi\right),
\label{eq:D}
\end{equation}

where $D=-1$ for a pure spin singlet~\cite{Afik:2020onf}. Near threshold the continuum itself is dominated by spin-singlet, maximally entangled pairs from $gg$ fusion~\cite{Kiyo:2008bv,Afik:2020onf}, and the bound state strengthens this further. Two observables built from $\varphi$ separate spin-singlet from spin-triplet configurations. The helicity observable is $c_{\rm hel}=\cos\varphi$. Its companion $c_{\rm han}$ is the cosine of the same angle after the sign of one lepton's momentum component along the top-quark direction is reversed. The experimental analyses combine these observables with $\mtt$~\cite{CMS:2025kzt,ATLAS:2026toponium}. The observable $c_{\rm han}$ was introduced in Ref.~\cite{Aguilar-Saavedra:2022uye}, and Ref.~\cite{Aguilar-Saavedra:2024mnm} proposed analyzing the lepton angles in bins of $\mtt$ to identify a toponium excess.

\subsection{A parity-sensitive observable}\label{sec:parity}

\begin{figure}[H]
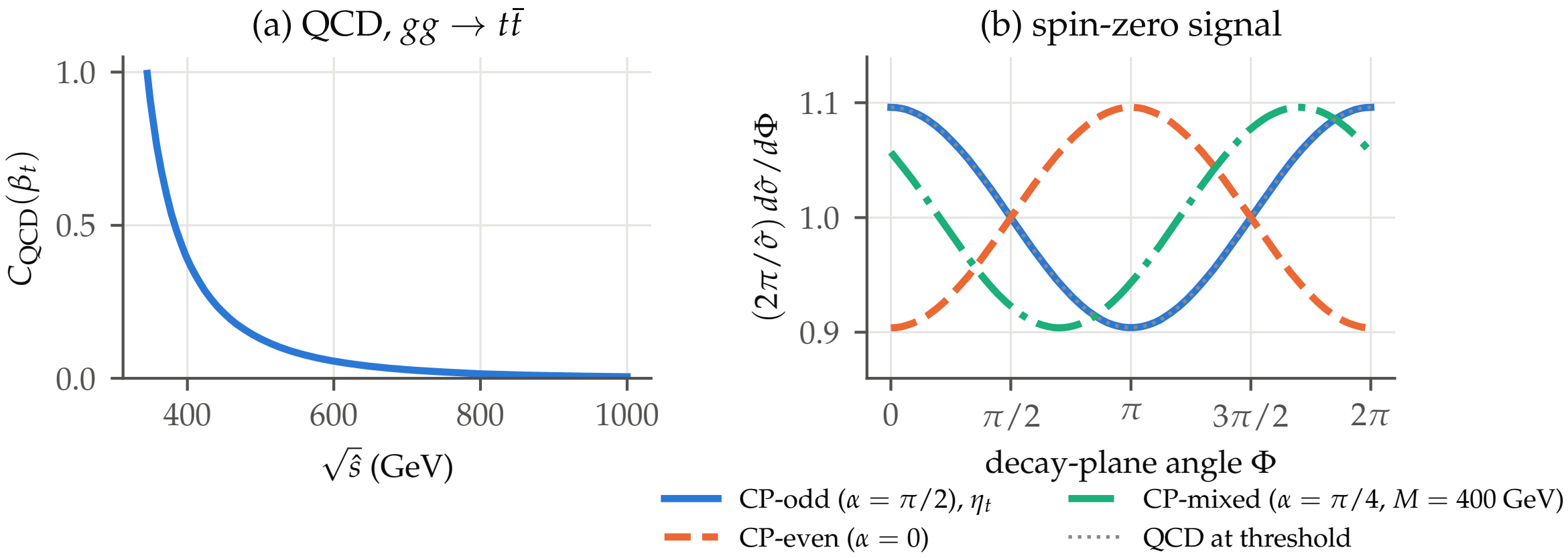

\begin{adjustwidth}{-\extralength}{0cm}
\centering
% ==== embedded figure fig_phi: generated by make_figures.py (matplotlib PGF backend); regenerate rather than edit ====
\begingroup%
\makeatletter%
%
\makeatother%
\endgroup%
% ==== end of embedded figure fig_phi ====
\caption{Decay-plane correlations in $\ttbar\to b\bar bW^+W^-$ at leading order~\cite{Barger:2011pu}. (\textbf{a}) The coefficient $C_{\rm QCD}(\beta_t)$ of Equation~(\ref{eq:Phi}) for $gg\to\ttbar$ as a function of the partonic energy. (\textbf{b}) Normalized distributions of the decay-plane angle $\Phi$, Equation~(\ref{eq:PhiH}), for a spin-zero state with CP-odd ($\alpha=\pi/2$), CP-even ($\alpha=0$), and CP-mixed ($\alpha=\pi/4$) couplings to the top quark ($\alpha$ is the coupling phase defined before Equation~(\ref{eq:PhiH})), with $k=0.096$ for $m_t=172.5$ GeV and $M_W=80.37$ GeV. The CP-mixed curve is evaluated at $M=400$ GeV. The QCD distribution at threshold (dotted) coincides with the CP-odd curve, which also describes the $\etat$, and the CP-even curve has the opposite phase.}
\label{fig:phi}
\end{adjustwidth}
\end{figure}

The azimuthal angle $\Phi$ between the $t$ and $\bar t$ decay planes, measured about the top-quark direction in the $\ttbar$ rest frame, is directly sensitive to parity. Each plane contains the top-quark direction and the momentum of the corresponding $W$ boson (or $b$ quark). For the partonic cross section $\hat\sigma$ of $gg\to\ttbar\to b\bar bW^+W^-$ at leading order in QCD~\cite{Barger:2011pu},

\begin{equation}
\frac{1}{\hat\sigma}\frac{d\hat\sigma}{d\Phi}=\frac{1}{2\pi}\left[1+C_{\rm QCD}(\beta_t)\,k\cos\Phi\right],\qquad k=\left(\frac{\pi}{4}\right)^2\left(\frac{1-2\rho_W}{1+2\rho_W}\right)^2,\quad \rho_W=\frac{M_W^2}{m_t^2},
\label{eq:Phi}
\end{equation}

with $\beta_t=v$ of Equation~(\ref{eq:v}) and $k\simeq 0.10$. The coefficient $C_{\rm QCD}$, given as a function of $\beta_t$ in Ref.~\cite{Barger:2011pu}, tends to 1 as $\beta_t\to0$ and to 0 as $\beta_t\to1$. The modulation is therefore largest just above threshold, where the spin singlet dominates, and has nearly disappeared by a partonic center-of-mass energy $\sqrt{\hat s}\approx800$ GeV (Figure~\ref{fig:phi}a).

\section{Experimental Observation}\label{sec:exp}

The first indications did not come from a dedicated toponium search. Near threshold, the data had disagreed with next-to-leading-order Monte Carlo predictions in several ATLAS and CMS analyses, including the observations of entanglement~\cite{ATLAS:2023fsd,CMS:2024pts} and the CMS search for heavy scalars and pseudoscalars decaying to $\ttbar$~\cite{CMS:2025AH}. CMS showed that adding a simplified $\etat$ contribution could remove the disagreement~\cite{CMS:2024pts,CMS:2025AH}, following the proposal of Ref.~\cite{Fuks:2021xje}.

\subsection{CMS dilepton channel}\label{sec:cmsdilepton}

Using 138 fb$^{-1}$ at $\sqrt s=13$ TeV, CMS fitted the reconstructed $\mtt$ distribution together with $c_{\rm hel}$ and $c_{\rm han}$. It observed an excess over fixed-order perturbative QCD near threshold with a significance above five standard deviations~\cite{CMS:2025kzt}. Interpreted with a simplified ${}^1S_0^{[1]}$ model, the excess corresponds to $\sigma(\etat)=8.8^{+1.2}_{-1.4}$ pb. CMS noted that the data do not exclude alternative explanations such as an additional pseudoscalar boson.

\subsection{ATLAS dilepton channel}\label{sec:atlasdilepton}

With 140 fb$^{-1}$, ATLAS sorted events into nine regions of $c_{\rm hel}$ and $c_{\rm han}$ and fitted $\mtt$ in each~\cite{ATLAS:2025conf,ATLAS:2026toponium}. The baseline model is perturbative QCD reweighted to next-to-next-to-leading order (NNLO) in QCD and next-to-leading order (NLO) in the electroweak coupling. The published analysis rejects it with an observed (expected) significance above $8\sigma$ ($6\sigma$). The nominal signal model is the Green's-function reweighting (GFRW) of Ref.~\cite{Fuks:2024yjj}. With it, the fitted cross section is $9.3^{+1.4}_{-1.3}$ pb, about 45\% above the reference normalization of 6.43 pb used for that model. The excess lies at low $\mtt$ and large $c_{\rm hel}$, as expected for a spin singlet. ATLAS also used the \textsc{bb4}$\ell$ generator as an alternative perturbative baseline. It computes $pp\to b\bar b\,\ell^+\nu\,\ell^-\bar\nu$ at NLO in QCD matched to parton showers, so it describes $\ttbar$ and $tW$ production together, with off-shell and interference effects~\cite{Jezo:2016ujg}. With this baseline, the result is $8.5^{+1.2}_{-1.1}$ pb. With the simplified $\etat$ model it is $13.1^{+1.9}_{-1.7}$ pb.

A second ATLAS analysis uses the normalized $e\mu$ invariant-mass and azimuthal-angle distributions in the same data set, without reconstructing the top quarks. The data are better described when a quasi-bound-state contribution is added, and fits to the $e\mu$ invariant mass find it with a significance above three standard deviations. With the GFRW model the signal strength is $1.22\pm0.38$, which corresponds to $7.8\pm2.4$ pb~\cite{ATLAS:2026qre}.

\subsection{CMS lepton+jets channel}\label{sec:cmsljets}

The lepton+jets analysis replaces $\mtt$, whose resolution near threshold is limited, by the relative velocity of the reconstructed top and antitop. It also uses parity-sensitive angular observables, which require identifying the down-type jet from the hadronic $W$ decay~\cite{CMS:2026lj}. The excess is observed at 6.1 standard deviations, with a cross section of $5.1\pm0.9$ pb in the simplified pseudoscalar model. The central value lies below the CMS dilepton result of 8.8 pb in the same model.

CMS also fits the nominal ATLAS Green's-function model, with the same reference normalization. It finds $2.8\pm0.5$ pb, i.e., a signal strength of 0.44, with an observed significance of 5.9 standard deviations. CMS attributes most of the difference from its $\etat$ result to the restriction of the Green's-function model to $\mtt<350$ GeV. About 62\% of the simulated $\etat$ events lie below this cutoff, and the corresponding $\etat$ contribution of about 3.2 pb agrees with the Green's-function value. CMS notes that both models are approximate~\cite{CMS:2026lj}.

\begin{table}[H]
\begin{adjustwidth}{-\extralength}{0cm}
\centering
\caption{Measurements of the $\ttbar$ threshold excess at $\sqrt s=13$ TeV. The quoted cross section is the excess over the perturbative-QCD baseline as defined in each analysis, and depends on the signal model (Section~\ref{sec:interp}). ``GFRW'' denotes the Green's-function-reweighted model of Ref.~\cite{Fuks:2024yjj}, which is defined only for $\mtt<350$ GeV, and ``$\etat$'' denotes the simplified pseudoscalar model of Refs.~\cite{Fuks:2021xje,Maltoni:2024csn}. The reference normalization, the cross section to which both templates are normalized, is not a prediction of the excess over a fixed-order baseline, which already contains part of the same threshold terms (Section~\ref{sec:missing} and Ref.~\cite{Nason:2026thr}).}
\label{tab:results}
%
\makeatother%
\endgroup%
% ==== end of embedded figure fig_measurements ====
\caption{Cross sections of the $\ttbar$ threshold excess from Table~\ref{tab:results}~\cite{CMS:2025kzt,ATLAS:2026toponium,CMS:2026lj,ATLAS:2026qre}. Circles denote the simplified $\etat$ model and squares the GFRW model. The two ATLAS dilepton points, and likewise the two CMS lepton+jets points, are fits to the same data with different signal templates, so their separation measures the model dependence, which includes the restriction of the GFRW template to $\mtt<350$ GeV. The ATLAS $e\mu$ point uses only lepton kinematics, in the data set of the ATLAS dilepton analysis. The dashed line is the reference normalization of 6.43 pb, for which no uncertainty is quoted, and the significances are listed at right.}
\label{fig:measurements}
\end{adjustwidth}
\end{figure}

\subsection{Comparison of the results}\label{sec:comparison}

Two experiments, in two decay channels, independently establish an excess at threshold with the angular signature of a spin singlet (Table~\ref{tab:results} and Figure~\ref{fig:measurements}). Results obtained with the same signal model differ between analyses. With the simplified $\etat$ model they range from 5.1 to 13.1 pb, and with GFRW from 2.8 to 9.3 pb. Within a single data set the extracted cross section also depends on the signal model ($9.3$ versus $13.1$ pb for ATLAS and $2.8$ versus $5.1$ pb for CMS lepton+jets).

\section{Modeling the Threshold}\label{sec:model}

\subsection{The simplified $\eta_t$ model}\label{sec:etamodel}

In 2021 Fuks, Hagiwara, Ma, and Zheng made a practical proposal for bringing toponium into LHC analyses~\cite{Fuks:2021xje}. The bound-state enhancement of color-singlet $gg\to\ttbar$ production is dominated by the ${}^1S_0^{[1]}$ state~\cite{Kiyo:2008bv}. Fuks \emph{et al.} showed that this contribution to the Run~2 dilepton data would be concentrated at small dilepton azimuthal separation and invariant mass, where ATLAS had reported an excess. As a simple event generator for this signal, they proposed a narrow pseudoscalar resonance $\etat$, with mass near $2m_t$ and effective couplings to gluons and to $\ttbar$.

This simplified $\etat$ model, also developed in Refs.~\cite{Maltoni:2024csn,Aguilar-Saavedra:2024mnm}, underlies the CMS observation. The model is simple to use, but a single Breit--Wigner resonance does not reproduce the continuum of S-wave states, the Sommerfeld-enhanced region above threshold, or the correlation between $\mtt$ and the top-quark momentum.

\subsection{Green's-function reweighting}\label{sec:gfrw}

Fuks, Hagiwara, Ma, and Zheng then developed a more complete description~\cite{Fuks:2024yjj}. The leading-order matrix elements for color-singlet S-wave $\ttbar$ production and decay are reweighted by the ratio of the interacting to the free momentum-space Green's function of Equation~(\ref{eq:green}). The Green's function is evaluated with the QCD potential in Coulomb gauge. Because the reweighting acts on the full $bW^+\bar bW^-$ matrix elements, the spin correlations and off-shell kinematics of the decay products are kept, and the events can be showered with standard tools.

The method is applied only where the nonrelativistic description is valid, i.e., for $\mtt<350$ GeV and top-quark momentum below 50 GeV in the pair rest frame. This model is labeled ``GFRW'' in Table~\ref{tab:results}. The prospects in the lepton+jets channel were later studied with the same approach~\cite{Fuks:2025single}, and the two models are compared in Ref.~\cite{Fuks:2025moriond}.

\subsection{Interpretation of the cross section}\label{sec:interp}

The extracted cross section is model dependent because it is defined relative to a perturbative baseline and to a signal template. Different templates distribute the same excess differently in $\mtt$ and in the angular variables. ATLAS finds that the simplified $\etat$ model predicts a smaller excess than GFRW in the lowest $\mtt$ bin, so a larger normalization is needed to fit the data~\cite{ATLAS:2026toponium}. The reference normalization of 6.43 pb itself depends on the $\mtt$ integration range and carries no stated uncertainty~\cite{ATLAS:2026toponium}.

Measured cross sections should therefore be compared with each other and with theory only at matched definitions. The significance of the excess depends less on the signal model than the cross section does (Table~\ref{tab:results}). The difference between the CMS dilepton and lepton+jets values may also reflect, in part, the different observables and acceptances. It is too early to say whether any remaining difference is physical.

\subsection{Missing ingredients}\label{sec:missing}

Several ingredients are missing from the current signal models, and each affects the comparison with data.
\begin{itemize}
\item \emph{Matching to fixed order.} The resummed threshold contribution must be combined with the NNLO perturbative continuum without double counting. ATLAS assumed minimal overlap between GFRW and the baseline~\cite{ATLAS:2026toponium}, although the NLO and NNLO calculations already contain, respectively, the first order and the first two orders of the $(\alpha_s/v)^n$ series that GFRW resums~\cite{Nason:2026thr}. Perturbative QCD itself, with off-shell effects and higher orders in $\alpha_s/v$, produces much of the threshold enhancement. For $\mtt<350$ GeV the resummed cross section differs from its truncation at third order in $\alpha_s/v$ by less than 1 pb. In the same region Nason \emph{et al.} estimate the excess over NLO generators at about 4 pb and expect about 2 pb less over an NNLO generator. They also provide public NLO generators matched to parton showers (NLO+PS) that add the resummed threshold corrections to NLO QCD, one of them with full off-shell effects, for use in experimental analyses~\cite{Nason:2026thr}. The ATLAS results, 9.3 pb with the nominal baseline reweighted to NNLO and 8.5 pb with \textsc{bb4}$\ell$, do not isolate this difference, because \textsc{bb4}$\ell$ is NLO in its kinematic distributions but also includes off-shell and $tW$ interference effects~\cite{ATLAS:2026toponium}.
\item \emph{Higher-order potential and uncertainties.} The NLO predictions of Section~\ref{sec:green} now come with a systematic study of their uncertainties that includes the color-singlet and color-octet Green's functions~\cite{Garzelli:2024uhe,Garzelli:2026thr}. The Green's function also depends on $m_t$, and this dependence has not yet been propagated in the experimental analysis~\cite{ATLAS:2026toponium}.
\item \emph{Beyond ${}^1S_0^{[1]}$.} P-wave and color-octet contributions, and quark-initiated channels, are absent from the nominal models~\cite{ATLAS:2026toponium}. A general-purpose implementation of threshold effects for singlet and octet states has been added to \textsc{Pythia}~\cite{Sjostrand:2025thr}, following Ref.~\cite{Fadin:1990wx}. It now includes angular correlations in the decay of a pseudoscalar $\ttbar$ state, though not the transition to uncorrelated decays above threshold~\cite{Sjostrand:2026rcz}.
\end{itemize}

None of these open issues calls the observation into question. Every perturbative baseline considered by ATLAS fails to describe the data without a threshold component. The goodness-of-fit probability, computed with a saturated model, changes from 0.98 with the GFRW component to below $10^{-5}$ without it~\cite{ATLAS:2026toponium}. Statistical analyses outside the collaborations reach the same conclusion~\cite{Fuks:2025hks}.

\section{Implications}\label{sec:implications}

\subsection{QCD in a new regime}\label{sec:qcdregime}

Toponium is the smallest and heaviest QCD bound state. Because $\Gamma_t\gg\Lambda_{\rm QCD}$, it probes the strong interaction in a regime where perturbation theory applies. The angular structure of the LHC excess agrees with the spin-singlet prediction of the Coulomb-resummed Green's function of Equation~(\ref{eq:green}), whose finite-width form was introduced by Fadin and Khoze~\cite{Fadin:1987wz} and developed over the following decades~\cite{Fadin:1990wx,Sumino:1992ai,Hagiwara:2008df,Kiyo:2008bv,Sumino:2010bv}. At the present $\mtt$ resolution, the first three orders of the $(\alpha_s/v)^n$ series reproduce most of the resummed enhancement~\cite{Nason:2026thr}, so the data do not yet isolate the resummation itself. The NLO calculation contains the first of these orders and the NNLO calculation the first two. The third is available only in the threshold approximation. A precise comparison of the rate awaits the theoretical improvements of Section~\ref{sec:missing}.

The threshold enhancement is small compared with the total $\ttbar$ rate, about 1\% of the selected dilepton events in the ATLAS signal regions~\cite{ATLAS:2026toponium}. It must nevertheless be included in any precision description of top-pair production at low $\mtt$.

\subsection{Spin correlations and entanglement}\label{sec:entanglement}

The observations of entanglement between top and antitop near threshold~\cite{ATLAS:2023fsd,CMS:2024pts} use the same region and the same observable $D$ of Equation~(\ref{eq:D}). ATLAS measures $D=-0.537\pm0.002\,({\rm stat})\pm0.019\,({\rm syst})$ for $340<\mtt<380$ GeV, more than five standard deviations beyond the limit $D=-1/3$ for separable states~\cite{ATLAS:2023fsd}, and CMS observes entanglement at 5.1 standard deviations~\cite{CMS:2024pts}. CMS has also observed entanglement at high $\mtt$ in lepton+jets events, where the pair is produced mainly in a spin-triplet state~\cite{CMS:2024zkc}.

The bound-state contribution is a pure spin singlet and increases the measured $|D|$. The ATLAS value is more negative than its baseline prediction, $-0.470\pm0.017$, from a generator without bound-state effects. Reweighting the $\mtt$ lineshape for these effects changed the result by only 0.5\%~\cite{ATLAS:2023fsd}, but such a reweighting misses the direct spin-singlet contribution to $D$. CMS included an $\etat$ component in its prediction (Section~\ref{sec:exp})~\cite{CMS:2024pts}. Predictions for entanglement and Bell-type measurements near threshold must therefore include the threshold-enhanced spin-singlet contribution, whether modeled as a toponium component or through higher orders in $\alpha_s/v$. Reference~\cite{Nason:2025hix} finds that the latter substantially reduces the tension with the measured spin correlations. Conversely, the spin observables help separate the bound-state contribution from the continuum.

An earlier ATLAS measurement of the dilepton azimuthal correlation found a spin-correlation strength 1.25 times the NLO QCD expectation, a 2.2 standard-deviation excess. The excess is not concentrated near threshold, and one fixed-order NLO calculation is consistent with the data~\cite{ATLAS:2019zrq}, so it does not point to a threshold effect.

\subsection{Bell nonlocality and quantum information}\label{sec:bell}

Within quantum mechanics, Bell nonlocality is a stronger property of a state than entanglement, and the $\ttbar$ spin density matrix reconstructed from LHC data has not yet been shown to have it. A two-particle spin state is Bell nonlocal if suitably chosen spin measurements on it would violate a Bell inequality. Fabbrichesi \emph{et al.} evaluated the corresponding combination of $\ttbar$ spin correlations and found that a violation would be within reach of LHC data~\cite{Fabbrichesi:2021npl}, whereas Severi \emph{et al.}, including unfolding, concluded that it will be challenging to establish at the High-Luminosity LHC (HL-LHC)~\cite{Severi:2021cnj}. Observables designed for the threshold region improve the sensitivity~\cite{Aguilar-Saavedra:2022uye}.

Aguilar-Saavedra and Rodr\'iguez-Ben\'itez estimate that, with current LHC data, quantum steering would be observable at about 10 standard deviations and Bell nonlocality at about 9, if the systematic uncertainties can be controlled~\cite{Aguilar-Saavedra:2026grl}. From the CMS measurement of the spin density matrix, Afik \emph{et al.} find evidence for steerability at more than three standard deviations and no Bell correlation in the phase space probed so far~\cite{Afik:2026pxv}. Combining spin and quantum-information observables also improves the sensitivity to the toponium contribution itself~\cite{Antozzi:2026vdi}.

None of these measurements is a test of locality, i.e., a test of quantum mechanics against local hidden-variable theories, which is the purpose of a Bell test. The measured quantities are the momenta of the decay products. All components of these momenta commute, so a local hidden-variable theory can be constructed that reproduces the measured distributions exactly and, as a local theory, satisfies Bell's inequality~\cite{Abel:1992kz,Abel:2025locality}. The spin correlations are obtained from the angular correlations of the decay products only by assuming the quantum-mechanical description of the decay, and the relation between the two, in quantum mechanics and in local hidden-variable theories, is analyzed for top-quark pairs in Refs.~\cite{Abel:2025locality,Bechtle:2025appraisal}. Bell nonlocality and steering, as used above, are therefore properties of the spin density matrix reconstructed under that assumption. The same holds for the entanglement criterion $D<-1/3$ of Section~\ref{sec:entanglement}. It establishes entanglement given the quantum-mechanical description of top-quark decay, with its Standard Model spin-analyzing power, and not unconditionally~\cite{Abel:2025locality,Bechtle:2025appraisal}. The observables remain tests of the Standard Model prediction for the $\ttbar$ spin state, including its threshold-enhanced spin-singlet component, and they are used in that sense in this article.

Quantum-information observables also offer complementary sensitivity to new physics in the same region~\cite{Maltoni:2024csn}. At an $e^+e^-$ collider the pair is produced near threshold through the vector current, in the spin-triplet $\psi_t$ configuration (Section~\ref{sec:leptoncolliders}). For $e^+e^-$ collisions the spin correlations, entanglement, and Bell observables have been worked out in the Standard Model and the Standard Model effective field theory (SMEFT)~\cite{Maltoni:2024lep}. Pei, Li, and collaborators have formulated general entanglement criteria for multiparticle final states, which reduce to the Peres--Horodecki and Clauser--Horne--Shimony--Holt (CHSH) criteria for two fermions, and applied them to $\ttbar$, $tW$, $ttt$, and $\ttbar W$ production~\cite{Pei:2025non}.

\subsection{Top-quark mass and Yukawa coupling}\label{sec:massyukawa}

The ATLAS and CMS combination of Run~1 direct measurements gives $m_t^{\rm MC}=172.52\pm0.33$ GeV~\cite{ATLAS:2024dxp}. The relation of such direct measurements to the pole mass is under discussion, with estimates of the difference ranging from well below 1 GeV~\cite{Nason:2017cxd} to between 0.5 and 1 GeV~\cite{Hoang:2020iah,Butenschoen:2016lpz} (Section~\ref{sec:massmeas}). The mass matters well beyond the top sector, for example for the stability of the electroweak vacuum (Section~\ref{sec:vacuum}). The position and shape of the threshold enhancement depend on $m_t$ in a short-distance scheme, so the LHC threshold region offers an additional handle on the mass, weak at present because of the $\mtt$ resolution. A mass fit to the threshold region requires the same mass to be used in the Green's function and in the perturbative baseline.

The top Yukawa coupling enters threshold production through Higgs exchange in the $\ttbar$ potential~\cite{Strassler:1990nw}. At a hadron collider this effect is small compared with the QCD uncertainties, because the range of Higgs exchange is short compared with the size of toponium~\cite{Bai:2025toponia}. The more direct LHC probes are $\ttbar H$ production~\cite{CMS:2018uxb,ATLAS:2018mme} and $\ttbar$ kinematic distributions~\cite{CMS:2019art,ATLAS:2025yt}. In an $e^+e^-$ threshold scan, the Yukawa dependence of the cross section becomes measurable (Section~\ref{sec:leptoncolliders}).

\subsection{A heavy pseudoscalar near threshold}\label{sec:heavyA}

The CMS threshold excess appeared in its search for heavy Higgs bosons decaying to $\ttbar$~\cite{CMS:2025AH}. The corresponding ATLAS search tested masses of 400 GeV and above and found no significant deviation from the Standard Model~\cite{ATLAS:2024vxm}. A heavy pseudoscalar $A$ coupled to top quarks remains a possible contributor~\cite{Djouadi:2024lyv}, although conventional two-Higgs-doublet models alone cannot accommodate one that explains the excess~\cite{Lu:2024twj}. A spin-zero boson produced in $gg\to A\to\ttbar$ interferes with the QCD continuum and produces a characteristic peak--dip structure in $\mtt$~\cite{Dicus:1994bm}. Toponium, by contrast, is a feature of the continuum itself.

The two effects are hard to separate when $M_A$ is close to $2m_t$~\cite{CMS:2025AH}. Flacke \emph{et al.} showed that the bound-state enhancement must be applied to the new-physics amplitude and to its interference with the Standard Model, not only to the Standard Model amplitude~\cite{Flacke:2025tos}. Doing so reshapes the allowed region of the pseudoscalar mass and coupling, and the interference changes sign near $M_A\simeq 2m_t$. They conclude that Standard Model threshold effects can account for a sizable part of the measured enhancement, while current data remain compatible with an additional pseudoscalar. A proposal in which an elementary state mixes with the bound-state field has also been discussed~\cite{Matsuoka:2025mix}.

A light scalar that couples to the top quark, for example through mixing with the Higgs boson, adds an attractive Yukawa potential between the top and antitop, which deepens the binding and raises the $\etat$ production rate. The measured rate then excludes parameter space that the bound on Higgs decays to pairs of the light scalar leaves open~\cite{Bai:2025toponia}.

\subsection{Distinguishing toponium from a new pseudoscalar}\label{sec:distinguish}

The decay-plane correlation of Equation~(\ref{eq:Phi}) distinguishes CP-even from CP-odd couplings of a new state, but not by itself the $\etat$ from a CP-odd $A$, because both produce the same modulation. Consider a spin-zero boson $\phi$ coupled as $-(m_t/v_{\rm EW})\,\phi\,\bar t(aP_L+a^*P_R)t$, with $P_{L,R}=(1\mp\gamma_5)/2$, $a=|a|e^{i\alpha}$, and $v_{\rm EW}\simeq246$ GeV the Higgs vacuum expectation value (Section~\ref{sec:ewsb}). Its normalized signal distribution is~\cite{Barger:2011pu}

\begin{equation}
\frac{1}{\hat\sigma_\phi}\frac{d\hat\sigma_\phi}{d\Phi}=\frac{1}{2\pi}\left[1-k\cos(\Phi-2\chi)\right],\qquad \tan\chi=\frac{\tan\alpha}{\beta_t}.
\label{eq:PhiH}
\end{equation}

A CP-even coupling ($\alpha=0$, $\chi=0$) gives a modulation out of phase with the QCD distribution of Equation~(\ref{eq:Phi}). A CP-odd coupling ($\alpha=\pi/2$, $2\chi=\pi$) gives one in phase with it (Figure~\ref{fig:phi}b). The in-phase case applies to any ${}^1S_0$ source.

For intermediate $\alpha$ the modulation is shifted and acquires a $\sin\Phi$ component, a direct signal of CP violation. Because $\tan\chi$ grows as $1/\beta_t$, the CP-odd component controls the phase for states near threshold, where the CP-even amplitude is P-wave suppressed. Interference between the new-physics and QCD amplitudes changes both the rate and the $\Phi$ distribution~\cite{Barger:2011pu}.

If the excess is entirely Standard Model toponium, its spin-parity is ${}^1S_0$, its rate converges to the NRQCD prediction as the modeling improves, and no peak--dip structure appears above threshold. If a pseudoscalar that couples mainly to top quarks contributes, interference distorts the $\mtt$ spectrum away from $2m_t$, and the rate differs from the refined NRQCD prediction in a direction set by the sign of the interference. Data from the High-Luminosity LHC, analyzed with improved models of both possibilities, may distinguish them, although the peak--dip structure is hard to resolve at the present $\mtt$ resolution~\cite{Djouadi:2024lyv}.

In summary, the angular structure of the threshold excess agrees with the spin-singlet prediction of the Coulomb-resummed Green's function, although at the present $\mtt$ resolution the data do not yet isolate the resummation itself. Predictions for entanglement and Bell-type measurements near threshold must include the threshold-enhanced spin-singlet contribution, whether modeled as toponium or through higher orders in $\alpha_s/v$. Current data remain compatible with an additional pseudoscalar, and High-Luminosity LHC data analyzed with improved models of both possibilities may distinguish the two.

\section{The Top-Quark Mass and Width}\label{sec:props}

\subsection{Measurements of the mass}\label{sec:massmeas}

Direct mass measurements reconstruct the top quark from its decay products and calibrate the result against Monte Carlo simulation. The most precise single measurement, from CMS in the lepton+jets channel at 13 TeV, gives $m_t^{\rm MC}=171.77\pm0.37$ GeV~\cite{CMS:2023ebf}, and the Run~1 combination is quoted in Section~\ref{sec:massyukawa}. These measurements determine the top-quark mass parameter $m_t^{\rm MC}$ of the event generator used to build the templates. In the measurement of Ref.~\cite{CMS:2023ebf} the signal is simulated with the NLO+PS generator \textsc{Powheg}, in which the top mass is renormalized in the on-shell scheme at one loop, so the input parameter is the pole mass at that order~\cite{Frixione:2007nw}. How closely the fitted parameter tracks the pole mass once the parton shower, hadronization, and the leading-order decay are included remains under discussion. Nason argues that it should be read as the pole mass, with the inaccuracies of the generator propagated into its uncertainty as for any perturbative calculation~\cite{Nason:2017cxd}. Studies with NLO+PS generators of increasing accuracy estimate these uncertainties and find that they depend on the shower model used~\cite{FerrarioRavasio:2018whr}. Hoang and collaborators argue instead that the infrared cutoff of the shower turns the generator mass into a cutoff-dependent short-distance mass~\cite{Hoang:2018zrp,Hoang:2020iah}. In that approach, calibrating the \textsc{Pythia}~8 mass parameter against hadron-level QCD predictions for $e^+e^-$ collisions gave a difference from the pole mass of 0.6 to 0.9 GeV, depending on the order of the calculation, and agreement within uncertainties with the MSR mass at $R=1$ GeV~\cite{Butenschoen:2016lpz}. An update of the calibration for \textsc{Pythia}, \textsc{Herwig}, and \textsc{Sherpa} finds agreement with the MSR mass at $R=1$ GeV within 200 MeV and a difference from the pole mass of 350 to 600 MeV~\cite{Dehnadi:2023msm}. The MSR mass $m_t^{\rm MSR}(R)$ is a short-distance mass that depends on a scale $R$, which sets the infrared cutoff on the self-energy fluctuations absorbed into the mass~\cite{Hoang:2008yj,Hoang:2017suc}. It coincides with the $\overline{\rm MS}$ mass $m_t(m_t)$ at $R=m_t$, evolves to lower $R$ through a renormalization-group equation in $R$, and interpolates toward the pole mass as $R\to0$ while remaining free of the renormalon ambiguity of the pole mass. The choice $R=1$ GeV is close to the shower cutoff of the event generator, which is why, in this approach, $m_t^{\rm MC}$ is expected to lie near the MSR mass at that scale~\cite{Butenschoen:2016lpz}. CMS notes that the uncertainty in relating the direct measurement to a theoretically well-defined mass is currently comparable to its experimental uncertainty~\cite{CMS:2023ebf}.

The mass can also be extracted from cross sections, in a scheme fixed by the calculation. CMS obtained $m_t^{\rm pole}=170.5\pm0.8$ GeV from the triple-differential $\ttbar$ cross section, fitted together with $\alpha_s$ and the parton distributions~\cite{CMS:2019esx}. The extraction relies on the shape of the $\mtt$ distribution near threshold. CMS neglected Coulomb and soft-gluon resummation and estimated that including them would raise the extracted pole mass by about 1 GeV~\cite{CMS:2019esx}. The bound-state effects of Section~\ref{sec:threshold} are therefore part of the theoretical input for any mass extracted from the threshold region.

Ball, ter Hoeve, and Stegeman included toponium contributions in a global parton-distribution-function (PDF) analysis of differential $\ttbar$ data and found $m_t^{\rm pole}=172.81\pm0.26$ GeV at approximate next-to-next-to-next-to-leading order (N$^3$LO) in QCD with electroweak corrections~\cite{Ball:2026qno}. CMS has also extracted the $\overline{\rm MS}$ mass from the $\mtt$ dependence of the cross section and measured its running up to scales of about 1 TeV, consistent with the renormalization-group prediction~\cite{CMS:2019jul}. The Particle Data Group (PDG) averages of the extractions from inclusive and $\ttbar$+jet cross sections are $m_t^{\rm pole}=172.1\pm0.6$ GeV and $m_t(m_t)=162.5^{+2.1}_{-1.5}$ GeV in the $\overline{\rm MS}$ scheme of QCD, without the electroweak contributions included in the full-Standard-Model mass of Equation~(\ref{eq:yt})~\cite{ParticleDataGroup:2026mpi}. CPT invariance requires equal top and antitop masses, and CMS measures $m_t-m_{\bar t}=-0.15\pm0.19\,({\rm stat})\pm0.09\,({\rm syst})$ GeV~\cite{CMS:2016les}.

\subsection{The width}\label{sec:width}

The width lies far below the experimental resolution on the reconstructed top-quark mass. ATLAS nevertheless measured it directly from the shapes of the reconstructed decays, finding $\Gamma_t=1.76\pm0.33\,({\rm stat})^{+0.79}_{-0.68}\,({\rm syst})$ GeV~\cite{ATLAS:2017vgz}. An indirect determination combining the branching-fraction ratio $\mathcal B(t\to Wb)/\mathcal B(t\to Wq)$ ($q=d,s,b$) with the $t$-channel single-top cross section, gives $\Gamma_t=1.36\pm0.02\,({\rm stat})^{+0.14}_{-0.11}\,({\rm syst})$ GeV~\cite{CMS:2014mxl}. A preliminary ATLAS analysis at 13 TeV finds $\Gamma_t=1.9\pm0.5$ GeV~\cite{ATLAS:2019onj}. All agree with the Standard Model prediction at NNLO in QCD with NLO electroweak corrections, $\Gamma_t=1.33$ GeV for $m_t^{\rm pole}=172.5$ GeV~\cite{ParticleDataGroup:2026mpi}. A threshold scan at an $e^+e^-$ collider would measure $\Gamma_t$ to a few tens of MeV (Section~\ref{sec:leptoncolliders}).

\subsection{Comparison of mass schemes}\label{sec:schemes}

Figure~\ref{fig:masses} collects the direct measurements and the pole masses, together with the $\overline{\rm MS}$ mass $m_t(M_Z)$ of Equation~(\ref{eq:yt}). The differences between the schemes reflect their definitions. The four-loop QCD relation between the $\overline{\rm MS}$ and pole schemes raises $m_t(m_t)=162.5$ GeV by about 10 GeV, to a pole mass near 172 GeV, consistent with the pole-mass extractions and close to $m_t^{\rm MC}$~\cite{Marquard:2015qpa}. A comparison of a measured mass with theory is meaningful only within a stated scheme, and the comparison with the vacuum-stability bound in Figure~\ref{fig:masses} shows why the scheme matters.

\begin{figure}[H]
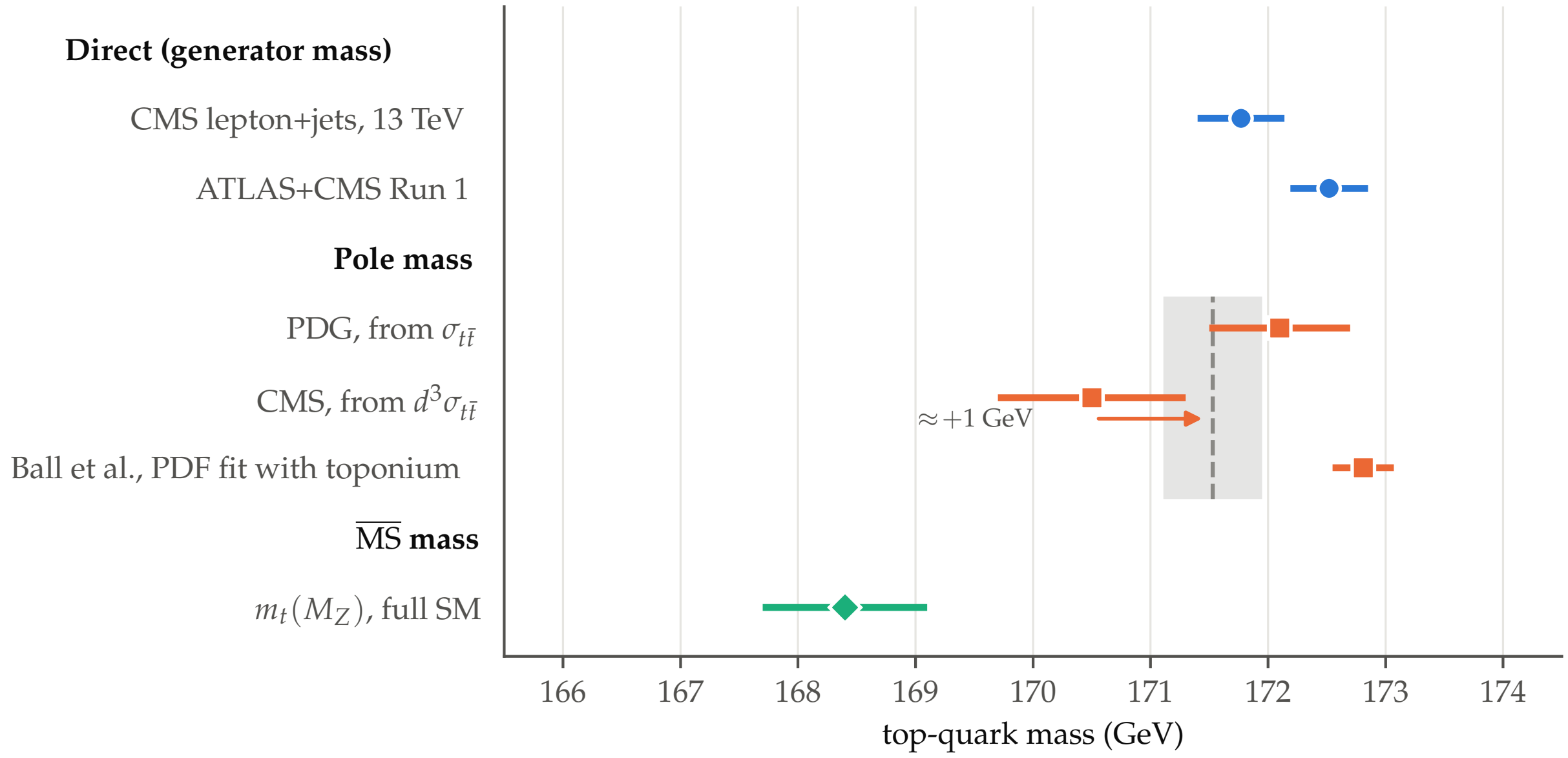

\begin{adjustwidth}{-\extralength}{0cm}
\centering
% ==== embedded figure fig_masses: generated by make_figures.py (matplotlib PGF backend); regenerate rather than edit ====
\begingroup%
\makeatletter%
%
\makeatother%
\endgroup%
% ==== end of embedded figure fig_masses ====
\caption{Determinations of the top-quark mass, grouped by scheme. The direct measurements, which determine the generator mass $m_t^{\rm MC}$, are the CMS lepton+jets measurement~\cite{CMS:2023ebf} and the ATLAS and CMS Run~1 combination~\cite{ATLAS:2024dxp}. The pole masses are the PDG average of extractions from inclusive and $\ttbar$+jet cross sections~\cite{ParticleDataGroup:2026mpi}, the CMS value from the triple-differential cross section~\cite{CMS:2019esx}, and the value of Ball \emph{et al.} from a global PDF fit that includes toponium~\cite{Ball:2026qno}. The short arrow shows the estimated shift of about $+1$ GeV of the CMS value from Coulomb and soft-gluon resummation. The $\overline{\rm MS}$ mass is $m_t(M_Z)=y_t(M_Z)\,v_{\rm EW}/\sqrt2$ of Equation~(\ref{eq:yt}) in the full Standard Model, from the Yukawa coupling of Antusch \emph{et al.}~\cite{Antusch:2025fpm} (see also Ref.~\cite{Huang:2020hdv}). It corresponds to a pole mass of 172.4 GeV. The gray band is the pole-mass bound for absolute stability of the electroweak vacuum in the Standard Model, assuming no new physics below the Planck scale, $171.53\pm0.42$ GeV~\cite{Buttazzo:2013uya}. It is a bound on the pole mass.}
\label{fig:masses}
\end{adjustwidth}
\end{figure}

In summary, direct measurements determine the mass parameter of the event generator, which in NLO+PS generators is the pole mass at one loop in the hard process. Its relation to the pole mass beyond that order is under discussion, and calibrations of the generator mass against QCD predictions find differences of 0.35 to 0.9 GeV. Masses extracted from the threshold region need the bound-state effects as theoretical input, and a comparison with theory, such as with the vacuum-stability bound, is meaningful only within a stated scheme. The measurements of the width agree with the Standard Model prediction.

\section{Top-Quark Production and Couplings}\label{sec:prodcoup}

\subsection{Single-top production}\label{sec:singletop}

ATLAS has now observed $s$-channel single-top production in a single experiment, at 5.1 standard deviations, with a cross section of $10.8\pm1.5\,({\rm stat})\pm1.6\,({\rm syst})$ pb at 13 TeV~\cite{ATLAS:2026zgr}. Single top quarks are produced through the electroweak interaction in three channels, i.e., $t$-channel $W$ exchange, $s$-channel $W$ production, and associated $tW$ production (Figure~\ref{fig:singletop}a--c). Single-top production was first observed at the Tevatron in 2009~\cite{CDF:2009itk,D0:2009isq}, and the $s$-channel process was observed there in 2014 in a combination of CDF and D0 data~\cite{CDF:2014uma}. At the LHC, CMS observed $tW$ production at 8 TeV~\cite{CMS:2014fut}. Single-top production measures $|V_{tb}|$ directly (Section~\ref{sec:ckm}). Single top quarks produced through the left-handed $Wtb$ vertex are highly polarized, and the polarization can be measured from the decay products (Section~\ref{sec:history}). ATLAS has measured the full polarization vector in $t$-channel production, finding a largest component of $0.91\pm0.10$, and has used it to bound the $tWb$ dipole operator~\cite{ATLAS:2022vym}.

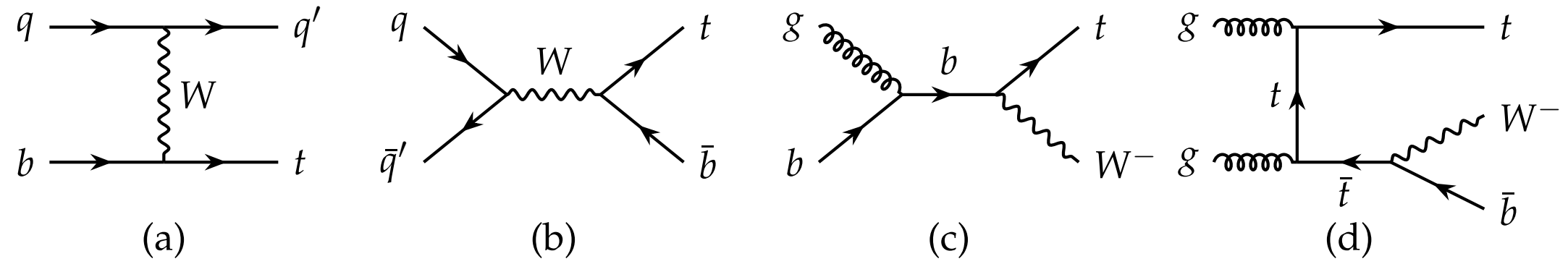
\begin{figure}[H]
\begin{adjustwidth}{-\extralength}{0cm}
\centering
\begin{tikzpicture}[scale=0.95]
% (a) t channel
\begin{scope}
\coordinate (v1) at (1.1,1.3); \coordinate (v2) at (1.1,0);
\draw[fermion] (0,1.3) -- (v1); \draw[fermion] (v1) -- (2.2,1.3);
\draw[fermion] (0,0) -- (v2); \draw[fermion] (v2) -- (2.2,0);
\draw[boson] (v1) -- (v2);
\node[lbl,left] at (0,1.3) {$q$}; \node[lbl,right] at (2.2,1.3) {$q'$};
\node[lbl,left] at (0,0) {$b$}; \node[lbl,right] at (2.2,0) {$t$};
\node[lbl,right] at (1.1,0.65) {$W$};
\node[lbl] at (1.1,-0.75) {(a)};
\end{scope}
% (b) s channel
\begin{scope}[xshift=3.6cm]
\coordinate (v1) at (0.8,0.65); \coordinate (v2) at (1.7,0.65);
\draw[fermion] (0,1.3) -- (v1); \draw[fermion] (v1) -- (0,0);
\draw[boson] (v1) -- (v2);
\draw[fermion] (v2) -- (2.5,1.3); \draw[fermion] (2.5,0) -- (v2);
\node[lbl,left] at (0,1.3) {$q$}; \node[lbl,left] at (0,0) {$\bar q'$};
\node[lbl,right] at (2.5,1.3) {$t$}; \node[lbl,right] at (2.5,0) {$\bar b$};
\node[lbl,above] at (1.25,0.7) {$W$};
\node[lbl] at (1.25,-0.75) {(b)};
\end{scope}
% (c) tW
\begin{scope}[xshift=7.4cm]
\coordinate (v1) at (0.8,0.65); \coordinate (v2) at (1.7,0.65);
\draw[gluon] (0,1.3) -- (v1); \draw[fermion] (0,0) -- (v1);
\draw[fermion] (v1) -- (v2);
\draw[fermion] (v2) -- (2.5,1.3); \draw[boson] (v2) -- (2.5,0);
\node[lbl,left] at (0,1.3) {$g$}; \node[lbl,left] at (0,0) {$b$};
\node[lbl,right] at (2.5,1.3) {$t$}; \node[lbl,right] at (2.5,0) {$W^-$};
\node[lbl,above] at (1.25,0.7) {$b$};
\node[lbl] at (1.25,-0.75) {(c)};
\end{scope}
% (d) doubly resonant
\begin{scope}[xshift=11.2cm]
\coordinate (v1) at (0.8,1.3); \coordinate (v2) at (0.8,0); \coordinate (v3) at (1.7,0);
\draw[gluon] (0,1.3) -- (v1); \draw[gluon] (0,0) -- (v2);
\draw[fermion] (v1) -- (2.6,1.3);
\draw[fermion] (v2) -- (v1);
\draw[fermion] (v3) -- (v2);
\draw[boson] (v3) -- (2.6,0.45); \draw[fermion] (2.6,-0.45) -- (v3);
\node[lbl,left] at (0,1.3) {$g$}; \node[lbl,left] at (0,0) {$g$};
\node[lbl,right] at (2.6,1.3) {$t$}; \node[lbl,right] at (2.6,0.45) {$W^-$}; \node[lbl,right] at (2.6,-0.45) {$\bar b$};
\node[lbl,left] at (0.8,0.65) {$t$}; \node[lbl,below] at (1.25,0) {$\bar t$};
\node[lbl] at (1.3,-0.75) {(d)};
\end{scope}
\end{tikzpicture}
\caption{Representative leading-order diagrams for single-top production in (\textbf{a}) the $t$ channel, (\textbf{b}) the $s$ channel, and (\textbf{c}) the $tW$ channel. (\textbf{d}) A doubly resonant diagram, $gg\to\ttbar\to tW^-\bar b$, which has the same final state as $tW$ production with an additional $\bar b$ quark, a next-to-leading-order correction to (\textbf{c}).}
\label{fig:singletop}
\end{adjustwidth}
\end{figure}

Beyond leading order, $tW$ production with an additional $b$ quark overlaps with $\ttbar$ production followed by top decay (Figure~\ref{fig:singletop}d). The two cannot be separated without a prescription, such as the diagram-removal and diagram-subtraction schemes~\cite{Frixione:2008yi}. ATLAS has measured a region where the interference between singly and doubly resonant production is large, and predictions that include the interference throughout describe the data best~\cite{ATLAS:2018ivx}. The off-shell \textsc{bb4}$\ell$ generator, used as an alternative baseline in the ATLAS toponium analysis, is of this kind~\cite{Jezo:2016ujg}. The treatment of the $tW$ interference therefore enters the definition of the threshold excess (Section~\ref{sec:missing}).

\subsection{Electroweak couplings}\label{sec:ewcouplings}

The couplings of the top quark to the $W$ and $Z$ bosons and the photon are fixed in the Standard Model, and the LHC tests them directly. In top decay, the helicity of the $W$ boson reflects the $V-A$ structure of the $Wtb$ vertex. At leading order the longitudinal fraction is $F_0=m_t^2/(m_t^2+2M_W^2)\simeq0.70$, the left-handed fraction is $F_L=1-F_0$, and the right-handed fraction $F_R$ vanishes for a massless $b$ quark. The ATLAS and CMS combination at 8 TeV gives $F_0=0.693\pm0.014$, $F_L=0.315\pm0.011$, and $F_R=-0.008\pm0.007$, in agreement with the Standard Model at NNLO~\cite{CMS:2020ezf}. Such measurements constrain anomalous $Wtb$ couplings.

The neutral-current couplings are probed in associated production. ATLAS measures $\sigma(\ttbar Z)=0.86\pm0.04\,({\rm stat})\pm0.04\,({\rm syst})$ pb at 13 TeV~\cite{ATLAS:2023eld}, and CMS measures $0.95\pm0.05\,({\rm stat})\pm0.06\,({\rm syst})$ pb~\cite{CMS:2019too}, both in agreement with the Standard Model. Single-top production with a $Z$ boson has been observed by both experiments~\cite{CMS:2018sgc,ATLAS:2020bhu}, CMS reported evidence for $tWZ$ production~\cite{CMS:2023krq} and has observed it at 5.8 standard deviations with data at 13 and 13.6 TeV~\cite{CMS:2025tre}, and $\ttbar\gamma$ production has been measured differentially~\cite{ATLAS:2024hmk}. ATLAS has also observed $\ttbar\gamma\gamma$ production, at 5.2 standard deviations~\cite{ATLAS:2025aps}. ATLAS measures the top-quark charge as $0.64\pm0.02\,({\rm stat})\pm0.08\,({\rm syst})$ in units of $e$, and D0 and ATLAS exclude an exotic quark of charge $-4/3$ in place of the top quark at more than five standard deviations~\cite{D0:2014oid,ATLAS:2013mkl}. Interpreted in the SMEFT, the $\ttbar Z$ and $\ttbar\gamma$ measurements constrain the operators that modify the $tZ$ and $t\gamma$ couplings~\cite{ATLAS:2023eld,ATLAS:2024hmk}.

\subsection{The charge asymmetry}\label{sec:asymmetry}

The forward-backward asymmetry measured at the Tevatron and the charge asymmetry measured at the LHC both agree with the NNLO Standard Model predictions. In QCD, top quarks produced in $q\bar q$ annihilation are emitted preferentially along the direction of the incoming quark. The effect arises at order $\alpha_s^3$ from interference between amplitudes of opposite charge-conjugation parity~\cite{Kuhn:1998jr}. At the Tevatron it appears as a forward-backward asymmetry $A_{FB}$. With its full data set, CDF measured $A_{FB}=0.164\pm0.047$~\cite{CDF:2012ctl}, above the next-to-leading-order prediction, and D0 measured $0.106\pm0.030$~\cite{D0:2014cda}. The CDF excess motivated many models with new vector bosons. One example is an asymmetric left-right model, proposed by Barger, Keung, and Yu, with a light $W'$ that couples the down quark to the top quark and a heavier $Z'$~\cite{Barger:2010mw}.

Models of a different kind added a $t$-channel contribution through a flavor-changing $Z'$ boson that couples the up quark to the top quark~\cite{Jung:2009jz}. Such a coupling also produces same-sign top pairs through $uu\to tt$, a process with top-quark number two that is negligible in the Standard Model. The couplings needed to explain the asymmetry implied copious same-sign top production at the LHC~\cite{Berger:2011ua}. CMS and ATLAS found no excess of same-sign top pairs in the early LHC data, which disfavored this interpretation~\cite{CMS:2011gff,ATLAS:2012iws}. Same-sign tops are the observable signature of transitions that change top-quark number by two. The analogue of $B$--$\bar B$ mixing, an oscillation between $t\bar q$ and $\bar tq$ mesons, does not occur, because the top quark decays before such mesons can form.

The Tevatron asymmetry was explained within the Standard Model. The complete NNLO QCD calculation, combined with the electroweak corrections, gives $A_{FB}=0.095\pm0.007$, an increase of about 27\% over the NLO QCD value~\cite{Czakon:2014xsa}. The final Tevatron combination, $A_{FB}=0.128\pm0.025$, agrees with it~\cite{CDF:2017cvy}.

At the LHC the symmetric $pp$ initial state turns the effect into a small difference between the rapidity distributions of top quarks and antiquarks, the charge asymmetry $A_C$. It is diluted because gluon fusion, the dominant production process, does not generate it. The ATLAS and CMS combination at 7 and 8 TeV agrees with the Standard Model~\cite{ATLAS:2017gkv}. With the full Run~2 data, ATLAS has found evidence for the asymmetry, $A_C=0.0068\pm0.0015$, which differs from zero by 4.7 standard deviations and agrees with the NNLO prediction~\cite{ATLAS:2022waa}. The differential measurements, as functions of $\mtt$ and of the transverse momentum and longitudinal boost of the $\ttbar$ system, also place bounds on several SMEFT coefficients~\cite{ATLAS:2022waa}.

\subsection{Top quarks at high momentum, in the forward region, and in nuclear collisions}\label{sec:boosted}

At transverse momenta well above $m_t$, the decay products of a hadronically decaying top quark merge into a single jet. Jet-substructure methods identify such boosted top quarks~\cite{Kaplan:2008ie}, now often with machine learning~\cite{Larkoski:2017jix}. They give access to the high-$\mtt$ tail, where heavy resonances and four-quark operators would appear. ATLAS, for example, has measured differential $\ttbar$ cross sections with a hadronically decaying top quark of transverse momentum above 355 GeV and used them to bound such contributions~\cite{ATLAS:2022xfj}. Measurements at high mass and at threshold thus constrain the same interactions from opposite ends of the spectrum. LHCb has observed top-quark production in the forward region, which is inaccessible to the central detectors and probes parton distributions at large momentum fraction~\cite{LHCb:2015nta,LHCb:2018usb}.

Top-quark pairs have also been observed in proton--lead~\cite{CMS:2017hnw,ATLAS:2024qdu} and lead--lead collisions~\cite{ATLAS:2024jji,CMS:2026hhb}, following first evidence in lead--lead collisions from CMS~\cite{CMS:2020aem}. These measurements probe the gluon distributions of nuclei. They can also be used to study the quark--gluon plasma. The top quark and the $W$ boson decay after mean times set by their lifetimes and boosts, so the hadronic $W$ decay products enter the plasma with a delay, and selecting the top-quark momentum varies the delay. Top quarks can probe the time structure of the plasma, although extensive studies require higher heavy-ion luminosities or a higher-energy collider~\cite{Apolinario:2017sob}.

\section{The Top-Quark Mass in Electroweak and Flavor Physics}\label{sec:ewsb}

The top quark is the only fermion whose coupling to the Higgs field is of order one,

\begin{equation}
y_t(M_Z)=\frac{\sqrt2\,m_t(M_Z)}{v_{\rm EW}}=0.967\pm0.004 .
\label{eq:yt}
\end{equation}

Here $y_t(M_Z)$ is the $\overline{\rm MS}$ Yukawa coupling at the renormalization scale $\mu=M_Z$ in the full Standard Model. The value is from Antusch, Hinze, and Saad, who evolve the 2024 PDG inputs, including the pole mass $m_t^{\rm pole}=172.4\pm0.7$ GeV, with two-loop electroweak and mixed QCD--electroweak running and matching~\cite{Antusch:2025fpm}. The corresponding running mass, defined as $m_t(\mu)\equiv y_t(\mu)\,v_{\rm EW}/\sqrt2$ with $v_{\rm EW}=(\sqrt2\,G_F)^{-1/2}\simeq246$ GeV and $G_F$ the Fermi constant, is $m_t(M_Z)=168.4\pm0.7$ GeV. Huang and Zhou find $168.3\pm0.8$ GeV with $v_{\rm EW}$ set to 246 GeV~\cite{Huang:2020hdv}. The running mass depends on this convention for the vacuum expectation value. With the running value $v(M_Z)=248.4$ GeV in place of $v_{\rm EW}$, the same coupling corresponds to 169.9 GeV~\cite{Antusch:2025fpm}. Because $y_t$ is of order one, the top quark tends to dominate wherever quantum corrections involve the Higgs sector or the breaking of flavor symmetry, and several sensitive tests of the Standard Model depend mainly on $m_t$.

\subsection{Electroweak precision}\label{sec:ewprecision}

The large mass splitting between the top and bottom quarks breaks custodial symmetry. It shifts the ratio of neutral- to charged-current strengths by~\cite{Veltman:1977kh}

\begin{equation}
\Delta\rho\simeq\frac{3\,G_F\,m_t^2}{8\sqrt2\,\pi^2}\simeq 0.0093 .
\label{eq:drho}
\end{equation}

Here $m_t\simeq172.5$ GeV is the pole mass. Using the running mass $m_t(M_Z)$ of Equation~(\ref{eq:yt}) instead gives $\Delta\rho\simeq0.0089$, and the difference is of higher order in $\alpha_s$ and the electroweak couplings. Because of this quadratic dependence, precision measurements at LEP and SLC constrained $m_t$ before the top quark was discovered~\cite{ALEPH:2005ab}. Today $m_t$ is a leading parametric input to the predicted $W$ mass, which shifts by about 6 MeV for each GeV of $m_t$~\cite{Awramik:2003rn}. Conversely, the global electroweak fit determines the pole mass indirectly as $175.2\pm1.8$ GeV, about 1.5 standard deviations above the ATLAS and CMS combination of direct measurements, $172.52\pm0.33$ GeV~\cite{ParticleDataGroup:2026mpi,ATLAS:2024dxp}.

The electroweak fit predicts $M_W=80353\pm6$ MeV, as quoted in Ref.~\cite{CMS:2024lrd}. The CMS measurement, $M_W=80360.2\pm9.9$ MeV~\cite{CMS:2024lrd}, agrees with it. The CDF value of $80433.5\pm9.4$ MeV~\cite{CDF:2022hxs} is in tension with both. With the slope quoted above~\cite{Awramik:2003rn}, the present $m_t$ uncertainty of a few hundred MeV contributes a few MeV to the prediction. An $e^+e^-$ threshold scan, which would determine $m_t$ to a few tens of MeV in a short-distance scheme (Section~\ref{sec:leptoncolliders}), would make this contribution negligible.

\subsection{The Higgs sector}\label{sec:higgssector}

Gluon fusion, the dominant Higgs production process at the LHC, proceeds mainly through a top-quark loop~\cite{Georgi:1977gs}. The top loop also contributes to $H\to\gamma\gamma$, where it interferes destructively with the $W$ loop. In Higgs-pair production through gluon fusion, the top-quark box and triangle diagrams interfere destructively, which makes the rate sensitive to the Higgs self-coupling~\cite{Glover:1987nx}.

The top loop also gives the largest correction to the Higgs mass parameter. For a cutoff $\Lambda$ this correction is of order $-3y_t^2\Lambda^2/(8\pi^2)$, so naturalness arguments focus on the top sector. In supersymmetry the top squarks cancel this correction (Section~\ref{sec:susyhiggs}).

Well before the discovery, the renormalization-group running of the top Yukawa coupling pointed to a heavy top quark. Pendleton and Ross found an infrared fixed point of the Yukawa coupling, which with the inputs of the time gave $m_t\approx135$ GeV~\cite{Pendleton:1980as}, and Hill showed that a Yukawa coupling that is large at the unification scale is focused toward a quasi-infrared fixed point, with $m_t\approx240$ GeV for the Standard Model~\cite{Hill:1980sq}. The fixed-point value is sensitive to $\alpha_s$ and lies above the measured mass. In the MSSM the same focusing gives $m_t\simeq(200~{\rm GeV})\sin\beta$, with two-loop running of the gauge and Yukawa couplings~\cite{Barger:1992ac}, so that the measured mass is reproduced for low $\tan\beta$; the fixed point then fixes the Higgs-sector predictions and the supersymmetric spectrum of radiative electroweak symmetry breaking~\cite{Barger:1993vu,Barger:1993gh,Bardeen:1993rv}.

The same quasi-fixed point governs the top-condensate model, in which the top quark is the source of electroweak symmetry breaking through a $\ttbar$ condensate~\cite{Bardeen:1989ds}. The minimal version predicted a top mass near 230 GeV for a cutoff scale near $10^{15}$ GeV, well above the measured value~\cite{Bardeen:1989ds}.

\subsection{Vacuum stability}\label{sec:vacuum}

At high scales the top Yukawa coupling drives the Higgs quartic coupling negative. In the Standard Model it crosses zero near $10^{10}$ GeV, and the electroweak vacuum is then metastable, with a lifetime much longer than the age of the universe. Absolute stability up to the Planck scale requires a pole mass $m_t<171.53\pm0.42$ GeV for the Higgs mass and $\alpha_s$ used in Ref.~\cite{Buttazzo:2013uya}. The PDG average and the CMS pole mass in Figure~\ref{fig:masses} lie within about one standard deviation of this bound, whereas the value of Ball \emph{et al.}~\cite{Ball:2026qno} lies about 2.6 standard deviations above it. Whether the vacuum is stable or metastable therefore depends on the mass determination and on its scheme ambiguity~\cite{Hoang:2020iah}. These statements assume that the Standard Model holds up to the Planck scale. New particles that couple to the Higgs boson at lower scales, such as the top squarks of Section~\ref{sec:susyhiggs}, can change the conclusion. The MSSM itself has no instability of this kind, because the quartic terms of its Higgs potential come from the $D$ terms and are fixed by the gauge couplings (Section~\ref{sec:susyhiggs}), so there is no free quartic coupling for the top Yukawa coupling to drive negative; large trilinear soft terms can instead produce charge- and color-breaking minima elsewhere in the scalar potential~\cite{Baer:2026ytu}.

\subsection{Rare processes}\label{sec:rare}

Because $x_t=m_t^2/M_W^2\simeq4.6$, the top-quark loop dominates flavor-changing neutral-current processes, in which the Glashow--Iliopoulos--Maiani (GIM) cancellation is broken by the up-type quark masses (Figure~\ref{fig:loops}). The Inami--Lim functions of $x_t$ govern the mixing of neutral $K$ and $B$ mesons and the rare decays $K\to\pi\nu\bar\nu$ and $K_L\to\mu^+\mu^-$~\cite{Inami:1980fz}. The ARGUS observation of $B^0$--$\bar B^0$ mixing in 1987~\cite{ARGUS:1987xtv} gave an early indication that the top quark was heavier than then expected. Its implications for $m_t$ were analyzed at once~\cite{Ellis:1987xg,Barger:1987xg}.

Top loops now control a set of well-measured rare processes. The $B_s\to\mu^+\mu^-$ branching ratio is predicted to be $(3.64\pm0.12)\times10^{-9}$ in the Standard Model and is measured at $(3.34\pm0.27)\times10^{-9}$~\cite{Czaja:2024the}. NA62 has observed $K^+\to\pi^+\nu\bar\nu$ with a branching ratio of $(13.0^{+3.3}_{-3.0})\times10^{-11}$~\cite{NA62:2024pjp}. The analysis of data through 2024 gives $(9.6^{+1.9}_{-1.8})\times10^{-11}$ with a significance above six standard deviations, consistent with Standard Model predictions near $8\times10^{-11}$~\cite{NA62:2026kpnn}. The indirect determinations of $V_{ts}$ and $V_{td}$ discussed in Section~\ref{sec:ckm} come from $B_d$ and $B_s$ mixing, which is also governed by top-quark loops.

\begin{figure}[H]
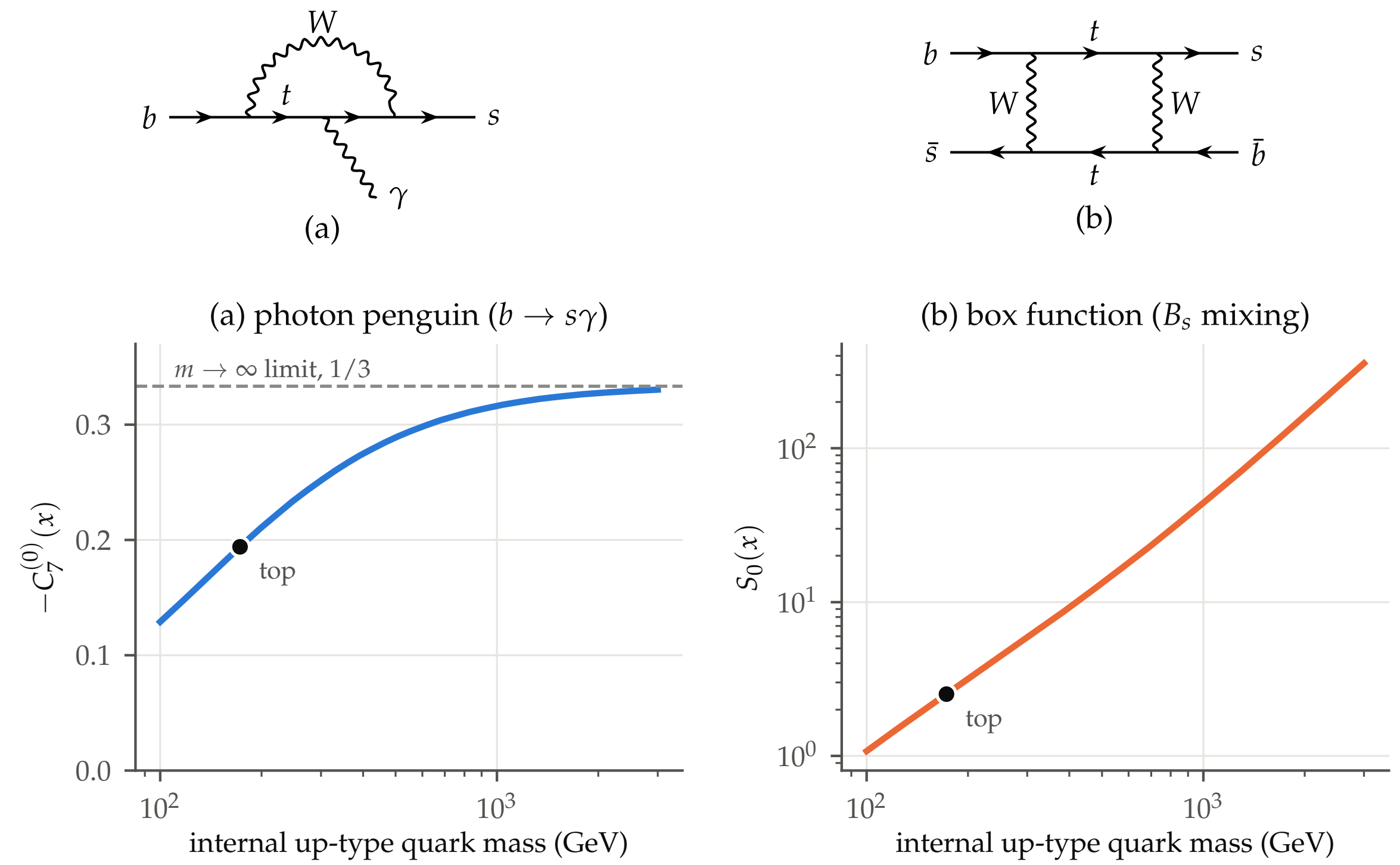

\begin{adjustwidth}{-\extralength}{0cm}
\centering
\begin{minipage}{0.48\textwidth}\centering
%
\makeatother%
\endgroup%
% ==== end of embedded figure fig_loops ====
\caption{Top row, representative one-loop diagrams in which a top quark and a $W$ boson circulate, (\textbf{a}) the photon penguin for $b\to s\gamma$, with the photon attached to the internal quark, and (\textbf{b}) a box diagram for $B_s$--$\bar B_s$ mixing. The full loop functions also include diagrams with the photon attached to the $W$ boson, Goldstone-boson exchange, and the crossed box. Bottom row, the leading-order loop functions~\cite{Inami:1980fz} versus the mass $m$ of the internal up-type quark, with $x=m^2/M_W^2$. (\textbf{a}) The photon-penguin function $C_7^{(0)}(x)$ that governs $b\to s\gamma$, which approaches $-1/3$ at large mass. (\textbf{b}) The box function $S_0(x)$ that governs $B_q$--$\bar B_q$ mixing, which grows as $x/4$ at large mass. The points mark the top quark. Both functions vanish as $x\to0$, so the top quark dominates over the lighter up-type quarks.}
\label{fig:loops}
\end{adjustwidth}
\end{figure}

\subsection{Flavor-changing top decays}\label{sec:fcnc}

LHC searches for flavor-changing neutral decays of the top quark itself, such as $t\to cH$, $t\to cZ$, and $t\to cg$, are sensitive at the level of $10^{-4}$. For example, combining searches in several Higgs-boson decay channels, ATLAS bounds $\mathcal B(t\to uH)<2.6\times10^{-4}$ and $\mathcal B(t\to cH)<3.4\times10^{-4}$ at 95\% confidence level (CL)~\cite{ATLAS:2023ujo,ATLAS:2024mih}, and CMS bounds them at $1.9\times10^{-4}$ and $3.7\times10^{-4}$~\cite{CMS:2024ubt}. The GIM mechanism also suppresses these decays, which involve loops of the light down-type quarks. The GIM suppression is then nearly complete, and the Standard Model branching ratios lie between about $10^{-15}$ and $10^{-11}$~\cite{Aguilar-Saavedra:2004mfd}.

Searches also bound charged-lepton-flavor-violating interactions, with limits below about $10^{-6}$ on $\mathcal B(t\to e\mu q)$~\cite{CMS:2022ztx,CMS:2023iul} and constraints on $\mu\tau qt$ interactions~\cite{ATLAS:2024njy}. A CMS search for baryon-number violation in top-quark production and decay improves earlier bounds by three to six orders of magnitude~\cite{CMS:2024dzv}. Because the Standard Model rates lie many orders of magnitude below the present sensitivity, any observation of such a decay would be a direct signal of new physics.

\section{The Top Yukawa Coupling and CP Violation}\label{sec:yukawa}

\subsection{Present constraints}\label{sec:cpconstraints}

The top Yukawa coupling is the largest coupling of the Higgs boson. Direct access at the LHC came with the observation of $\ttbar H$ production~\cite{CMS:2018uxb,ATLAS:2018mme}. The CP structure of the $Ht\bar t$ vertex has since been studied~\cite{Barger:2018tqn,Barger:2019ccj}. We write the coupling as

\begin{equation}
\mathcal L_{Ht\bar t}=-\frac{m_t}{v_{\rm EW}}\,g_{Htt}\,H\,\bar t\left(\cos\xi+i\gamma_5\sin\xi\right)t .
\label{eq:cpyukawa}
\end{equation}

The Standard Model corresponds to $g_{Htt}=1$ and $\xi=0$, and a nonzero $\xi$ signals CP violation.

The most recent ATLAS analysis of $t\bar tH$ and $tH$ production with $H\to\gamma\gamma$ combines 140 fb$^{-1}$ at 13 TeV with 164 fb$^{-1}$ at 13.6 TeV. It excludes $|\xi|>38^\circ$ at 95\% CL and excludes a purely CP-odd coupling at 5.8 standard deviations~\cite{ATLAS:2026cpgg}. In multilepton final states, CMS finds a CP-odd fraction of the $Ht\bar t$ coupling $|f_{CP}^{Htt}|=0.59$, with a 68\% confidence interval of 0.24 to 0.81. Combined with the $H\to ZZ$ and $H\to\gamma\gamma$ channels, the result is $|f_{CP}^{Htt}|<0.55$ at 68\% CL~\cite{CMS:2022cpml}.

\subsection{Interference as a probe of the phase}\label{sec:interference}

Processes in which the $Ht\bar t$ amplitude interferes with the $HWW$ amplitude are sensitive to the sign and phase of the top coupling relative to the gauge coupling, not only to its magnitude. For a heavy spin-zero state produced in $gg\to\ttbar$, the corresponding CP-sensitive observable is the phase of the decay-plane modulation in Equation~(\ref{eq:PhiH}).

Zheng and collaborators~\cite{Barger:2018tqn,Barger:2019ccj} have studied one process with this $Ht\bar t$--$HWW$ interference, single-top production in association with a Higgs boson through $t$-channel $W$ exchange. In the unitary gauge, the diagrams in which the Higgs is radiated from the top quark and from the $W$ boson interfere strongly and destructively, and their terms that grow with energy cancel in the Standard Model. In work by Barger, Hagiwara, and Zheng, this structure was shown to make $tHj$ production, where $j$ denotes a light-quark jet, a direct probe of the relative phase of the $Ht\bar t$ and $HWW$ couplings at the LHC~\cite{Barger:2018tqn,Barger:2019ccj}. The CP-odd observables are the azimuthal asymmetry between the $W$-emission and $tH$-production planes and the top-quark polarization normal to the production plane, both proportional to $\sin\xi$.

The CP properties of the top Yukawa coupling at an $e^+e^-$ collider were studied in Ref.~\cite{Hagiwara:2017ban}. In the SMEFT, the CP-violating phase arises from a dimension-six operator that also generates $\ttbar HH$ contact interactions. The resulting high-energy behavior of vector-boson-fusion amplitudes was analyzed in Ref.~\cite{Barger:2023wbg}.

\subsection{Muon colliders}\label{sec:muoncollider}

A high-energy muon collider can probe the same energy-growing interference terms. Cassidy, Dong, Kong, Lewis, Zhang, and Zheng studied $\ttbar H$, $tbH\mu\nu$, and $\ttbar H\nu\bar\nu$ production at multi-TeV muon colliders~\cite{Cassidy:2023lwd}. Above about 10 TeV, vector-boson fusion dominates. In the Standard Model the top-Yukawa and Higgs-gauge contributions to $\ttbar H\nu\bar\nu$ and $tbH\mu\nu$ combine so that terms growing with energy cancel. A CP-violating phase spoils this cancellation, so the cross sections are sensitive to $\xi$ despite projected statistical uncertainties of $O(50\%)$. An $e^+e^-$ threshold scan (Section~\ref{sec:leptoncolliders}) would constrain the magnitude of $y_t$, and these interference-based probes would constrain its phase.

\section{Top-Quark Flavor and Effective Field Theory}\label{sec:bsm}

\subsection{The third row of the CKM matrix}\label{sec:ckm}

Top-quark production and decay provide the only direct access to $|V_{tb}|$, $|V_{ts}|$, and $|V_{td}|$, which are otherwise inferred from loop processes. Table~\ref{tab:ckm} compares the two kinds of determination. Single-top measurements at the Tevatron and the LHC determine $|V_{tb}|$ to within about 2.5\%~\cite{ParticleDataGroup:2026mpi}. CMS has analyzed $t$-channel single-top production at 13 TeV, allowing for $s$- and $d$-quark contributions in both production and decay, and quotes results with and without the assumption of three-generation unitarity~\cite{CMS:2020vac}.

The global fit relies on top-quark loops, on Standard Model dynamics within those loops, and on three-generation unitarity~\cite{ParticleDataGroup:2026mpi}. Within it, $B_d$ and $B_s$ mixing with lattice-QCD input fixes the ratio $|V_{td}/V_{ts}|$. The direct bound on $|V_{td}|^2+|V_{ts}|^2$ is about thirty times the value implied by the fit, so top-quark measurements are not yet sensitive to $V_{ts}$ or $V_{td}$.

\begin{table}[H]
\centering
\caption{Direct and indirect determinations of the third row of the CKM matrix. The direct values come from single-top production~\cite{ParticleDataGroup:2026mpi,CMS:2020vac}, and the CMS limits are at 95\% CL. The global-fit values are from Ref.~\cite{ParticleDataGroup:2026mpi}, except $|V_{td}/V_{ts}|$, which is the determination from the $B_d$ and $B_s$ mixing frequencies with lattice-QCD input that enters the fit. The fit value of $|V_{td}|^2+|V_{ts}|^2$ is computed from the fitted $|V_{td}|$ and $|V_{ts}|$.}
\label{tab:ckm}
\small
\renewcommand{\arraystretch}{1.3}
\begin{tabular}{lcccc}
\toprule
 & Single-top & CMS, unitarity & CMS, no unitarity & \\
Quantity & average & assumed & assumed & Global fit \\
\midrule
$|V_{tb}|$ & $1.002\pm0.024$ & $>0.970$ & $0.988\pm0.024$ & fixed by unitarity \\
$|V_{td}|^2+|V_{ts}|^2$ & & $<0.057$ & $0.06\pm0.06$ & $1.8\times10^{-3}$ \\
$|V_{ts}|$ & & & & $0.04117^{+0.00079}_{-0.00068}$ \\
$|V_{td}|$ & & & & $0.00863^{+0.00019}_{-0.00017}$ \\
$|V_{td}/V_{ts}|$ & & & & $0.207\pm0.001\pm0.003$ \\
\bottomrule
\end{tabular}
\end{table}

The direct measurements are nonetheless the only determinations of the third row that assume neither three-generation unitarity nor Standard Model dynamics in loops, both of which new physics could alter. Measuring $|V_{ts}|$ in top decays would require identifying strange-quark jets from $t\to sW$ with high efficiency and purity, a capability being developed for future detectors~\cite{FCC:2025lpp}.

\subsection{The unitarity triangle and the CP phase}\label{sec:triangle}

The magnitude of $V_{td}$ also determines the CP-violating phase of the quark sector, through unitarity. Orthogonality of the first and third columns of the CKM matrix requires

\begin{equation}
V_{ud}V_{ub}^*+V_{cd}V_{cb}^*+V_{td}V_{tb}^*=0 .
\label{eq:ut}
\end{equation}

Divided by $V_{cd}V_{cb}^*$, Equation~(\ref{eq:ut}) describes a triangle in the complex plane with vertices at 0, 1, and

\begin{equation}
\bar\rho+i\bar\eta\equiv-\frac{V_{ud}V_{ub}^*}{V_{cd}V_{cb}^*} ,
\label{eq:rhoeta}
\end{equation}

which is independent of the phase convention~\cite{ParticleDataGroup:2026mpi}. In the Wolfenstein parameterization, with $\lambda=|V_{us}|\simeq0.225$, $A=|V_{cb}|/\lambda^2$, and $V_{ub}=A\lambda^3(\rho-i\eta)$ to leading order, one has $\bar\rho=\rho(1-\lambda^2/2+\dots)$ and $\bar\eta=\eta(1-\lambda^2/2+\dots)$. The two nontrivial sides of the triangle are

\begin{equation}
R_b=\left|\frac{V_{ud}V_{ub}^*}{V_{cd}V_{cb}^*}\right|\simeq\left(1-\frac{\lambda^2}{2}\right)\frac{1}{\lambda}\left|\frac{V_{ub}}{V_{cb}}\right|,\qquad
R_t=\left|\frac{V_{td}V_{tb}^*}{V_{cd}V_{cb}^*}\right|\simeq\frac{1}{\lambda}\left|\frac{V_{td}}{V_{ts}}\right|,
\label{eq:RbRt}
\end{equation}

A triangle is fixed by its three sides, so the magnitudes alone determine its angles, although not the sign of $\bar\eta$. The angle at the origin, $\gamma=\arg(-V_{ud}V_{ub}^*/V_{cd}V_{cb}^*)$, follows from

\begin{equation}
\cos\gamma=\frac{1+R_b^2-R_t^2}{2R_b},
\label{eq:cosgamma}
\end{equation}

and it is nearly equal to the phase $\delta$ of the standard parameterization. A triangle of nonzero area, i.e., $\bar\eta\neq0$, signals CP violation. All unitarity triangles have the same area, $J/2$, where $J\simeq A^2\lambda^6\bar\eta$ is the Jarlskog invariant~\cite{Jarlskog:1985ht}.

The top quark supplies the side $R_t$, because $|V_{td}/V_{ts}|$ comes from the ratio of the $B_d$ and $B_s$ mixing frequencies, i.e., from top-quark box diagrams. With $|V_{td}/V_{ts}|=0.207\pm0.003$, $|V_{ub}|=(3.89\pm0.16)\times10^{-3}$, $|V_{cb}|=(40.7\pm1.3)\times10^{-3}$, and $\lambda=0.22517$~\cite{ParticleDataGroup:2026mpi}, and including the $O(\lambda^2)$ correction to $R_t$, we find $R_t\simeq0.90$, $R_b\simeq0.41$, $\bar\eta\simeq0.37\pm0.02$, and $\gamma\simeq65^\circ\pm2^\circ$. The uncertainty on $\gamma$ comes almost entirely from $|V_{td}/V_{ts}|$. Because the third angle of the triangle is close to $90^\circ$, a change in $R_b$ moves the apex nearly along the direction of fixed $\gamma$, so $|V_{ub}|$ mainly limits $\bar\eta$ rather than $\gamma$.

This value of the phase uses no CP-violating measurement. It agrees with the direct determination from $B\to DK$ decays, $\gamma=(66.4^{+2.7}_{-2.8})^\circ$, and with the global-fit value $\delta=1.154\pm0.025$ rad, i.e., $66.1^\circ$~\cite{ParticleDataGroup:2026mpi}. The agreement is a test of the three-generation picture with top-quark loops. New physics in $B_q$ mixing would shift the value of $|V_{td}/V_{ts}|$ inferred from the mixing frequencies, and with it $R_t$, whereas the determination of $\gamma$ from tree-level decays would be largely unaffected. A fit that uses only tree-level inputs gives $\bar\eta=0.379^{+0.022}_{-0.021}$, compared with $\bar\eta=0.356\pm0.007$ from the full fit~\cite{ParticleDataGroup:2026mpi}.

\subsection{Effective field theory}\label{sec:eft}

Global fits in the SMEFT find no significant deviation from the Standard Model. The SMEFT provides a common language for combining top-quark measurements with Higgs, diboson, and electroweak precision data. Global fits now include the LHC Run~2 top-quark data~\cite{Brivio:2019ius,Ellis:2020unq,Ethier:2021bye,Celada:2024mcf}, most of them with SMEFT predictions at NLO in QCD~\cite{Brivio:2019ius,Ethier:2021bye,Celada:2024mcf}, and the experiments perform their own multi-operator analyses at detector level~\cite{CMS:2023xyc}.

The chromomagnetic dipole operator is constrained mainly by the $\ttbar$ rate and distributions~\cite{Ethier:2021bye}. Four-quark operators with two light and two heavy quarks are constrained by the $\ttbar$ distributions and the charge asymmetry~\cite{Ethier:2021bye}, and the electroweak top couplings by $\ttbar Z$, $tZq$, and $\ttbar\gamma$ (Section~\ref{sec:ewcouplings}). Four-heavy-quark operators are among the least constrained, because at tree level they are probed mainly by four-top and $\ttbar b\bar b$ production~\cite{DiNoi:2025fhq} (Section~\ref{sec:multitop}).

A fit that omits the bound-state contribution could attribute the excess at low $\mtt$ to operators that modify the $gg\to\ttbar$ vertex or the four-quark interactions. The threshold enhancement, whether modeled as an $\etat$ component or through higher orders in $\alpha_s/v$, must therefore be part of the Standard Model prediction. Conversely, the threshold region is itself sensitive to some operators. The CP-odd chromoelectric dipole moment of the top quark, for example, interferes with the Standard Model amplitudes for $\etat$ and continuum production, and angular asymmetries of the charged leptons near threshold can test it~\cite{Ma:2026qch}.

\section{The Top Quark and the Higgs Mass in Supersymmetry}\label{sec:susyhiggs}

In the minimal supersymmetric standard model (MSSM) the quartic coupling of the Higgs potential is fixed by the gauge couplings. At tree level this bounds the mass of the lighter CP-even Higgs boson by $m_h\le M_Z|\cos2\beta|$, where $\tan\beta$ is the ratio of the vacuum expectation values of the two Higgs doublets. The measured value $m_h\simeq125$ GeV therefore requires a large radiative correction, and the top quark supplies it. Because the top Yukawa coupling is of order one, Equation~(\ref{eq:yt}), the incomplete cancellation between top-quark and top-squark loops raises $m_h^2$ by an amount that grows as $m_t^4$~\cite{Haber:1990aw,Okada:1990vk,Ellis:1990nz}.

\subsection{The Higgs mass from top and top-squark loops}\label{sec:higgsmass}

When the two top squarks are heavy and their masses are not far apart, the leading one-loop result is~\cite{Carena:2002es,Draper:2011aa}

\begin{equation}
m_h^2\simeq M_Z^2\cos^22\beta+\frac{3\,m_t^4}{2\pi^2v_{\rm EW}^2}\left[\ln\frac{M_S^2}{m_t^2}+\frac{X_t^2}{M_S^2}\left(1-\frac{X_t^2}{12\,M_S^2}\right)\right].
\label{eq:mhsusy}
\end{equation}

Here $M_S^2=(m_{\tilde t_1}^2+m_{\tilde t_2}^2)/2$ is the mean squared top-squark mass, $X_t=A_t-\mu\cot\beta$ the top-squark mixing parameter, $A_t$ the top-squark trilinear coupling, and $\mu$ the higgsino mass parameter. In Equation~(\ref{eq:mhsusy}) $m_t$ is the $\overline{\rm MS}$ running mass, evaluated at $m_t$ or, after renormalization-group improvement, at $\sqrt{m_tM_S}$. Because the correction grows as $m_t^4$, using the pole mass instead overestimates $m_h$ by as much as 10 GeV~\cite{Haber:1996fp}.

Two features of Equation~(\ref{eq:mhsusy}) shape the phenomenology. The correction grows only logarithmically with the top-squark masses. For fixed $M_S$ it is largest at $|X_t|=\sqrt6\,M_S$, the maximal-mixing case. A Higgs mass of 125 GeV therefore requires either top squarks heavier than about 10 TeV or near-maximal mixing~\cite{Draper:2011aa}.

Equation~(\ref{eq:mhsusy}) is a leading-order guide and somewhat overestimates $m_h$~\cite{Carena:2002es}. Higher-order fixed-order calculations, combined with effective-field-theory resummation of the large logarithms, now give $m_h$ with a theoretical uncertainty of about 1 GeV in much of the parameter space. These calculations yield a slightly lower $m_h$ than earlier fixed-order results, implying a somewhat heavier supersymmetry scale~\cite{PardoVega:2015eno,Slavich:2020zjv}. The $m_t^4$ dependence also makes the top mass the input to which the prediction is most sensitive. For moderate to large $\tan\beta$ and TeV-scale top squarks, each GeV of variation in the top mass shifts the predicted $m_h$ by roughly one GeV, and the shift is larger at small $\tan\beta$~\cite{Slavich:2020zjv}. The relation between the mass measured at hadron colliders and a well-defined short-distance mass adds a further uncertainty, which an $e^+e^-$ threshold scan would remove (Section~\ref{sec:leptoncolliders})~\cite{Slavich:2020zjv}.

\subsection{Naturalness and electroweak fine-tuning}\label{sec:naturalness}

The top Yukawa coupling also controls how naturally the weak scale arises. Minimizing the one-loop Higgs potential gives~\cite{Baer:2012up,Baer:2012cf}

\begin{equation}
\frac{M_Z^2}{2}=\frac{m_{H_d}^2+\Sigma_d^d-\left(m_{H_u}^2+\Sigma_u^u\right)\tan^2\beta}{\tan^2\beta-1}-\mu^2\simeq-m_{H_u}^2-\Sigma_u^u-\mu^2 .
\label{eq:mzsusy}
\end{equation}

Here $m_{H_u}^2$ and $m_{H_d}^2$ are the soft-breaking Higgs mass parameters at the weak scale, with the various loop contributions to the one-loop effective potential contained in $\Sigma_u^u$ and $\Sigma_d^d$. The approximate form holds for moderate to large $\tan\beta$. The weak scale is natural when no term on the right-hand side is much larger than $M_Z^2/2$. Writing Equation~(\ref{eq:mzsusy}) as $M_Z^2/2=\sum_i C_i$, with

\begin{equation}
C_{H_d}=\frac{m_{H_d}^2}{\tan^2\beta-1},\qquad C_{H_u}=-\frac{m_{H_u}^2\tan^2\beta}{\tan^2\beta-1},\qquad C_\mu=-\mu^2,
\label{eq:Ci}
\end{equation}

and the terms $C_{\Sigma_d^d}$ and $C_{\Sigma_u^u}$ obtained from $C_{H_d}$ and $C_{H_u}$ by replacing $m_{H_d}^2$ and $m_{H_u}^2$ with $\Sigma_d^d$ and $\Sigma_u^u$, the electroweak fine-tuning measure is~\cite{Baer:2012up,Baer:2012cf}

\begin{equation}
\Delta_{\rm EW}=\frac{\max_i|C_i|}{M_Z^2/2}.
\label{eq:deltaEW}
\end{equation}

Two older measures are also in use. The log-derivative measure introduced by Ellis, Enqvist, Nanopoulos, and Zwirner and by Barbieri and Giudice,

\begin{equation}
\Delta_{p_i}=\max_i\left|\frac{\partial\log M_Z^2}{\partial\log p_i}\right|,
\label{eq:deltaBG}
\end{equation}

measures the sensitivity of $M_Z^2$ to the fundamental parameters $p_i$ of the theory~\cite{Ellis:1986yg,Barbieri:1987fn}. The high-scale measure

\begin{equation}
\Delta_{\rm HS}\sim\frac{\delta m_{H_u}^2}{m_h^2}
\label{eq:deltaHS}
\end{equation}

compares the Higgs mass with the correction $\delta m_{H_u}^2$ to $m_{H_u}^2$ from running between the high and weak scales~\cite{Murayama:2000dw,Kitano:2006gv,Papucci:2011wy}. With equal soft-breaking masses $\tilde m$ for the left- and right-handed top squarks, the top-squark loop gives~\cite{Murayama:2000dw}

\begin{equation}
\delta m_{H_u}^2\simeq-\frac{3y_t^2}{8\pi^2}\,\tilde m^2\ln\frac{\Lambda^2}{\tilde m^2},
\label{eq:dmhu}
\end{equation}

where $\Lambda$ is the high scale at which the running starts. Because the logarithm is large, a small $\Delta_{\rm HS}$ requires light top squarks. Estimated with these two measures, naturalness seemed to require three third-generation squarks below 1 TeV~\cite{Baer:2025hllhc}.

Baer and collaborators argue that both measures overestimate the fine-tuning, by factors of 10 to $10^3$~\cite{Baer:2025hllhc,Baer:2025uzx}. The measure $\Delta_{p_i}$ neglects correlations among the soft parameters, which in a complete model are computable in terms of a few quantities such as the gravitino mass. The measure $\Delta_{\rm HS}$ neglects that the running of $m_{H_u}^2$ depends on $m_{H_u}^2$ itself and must be large enough to break electroweak symmetry radiatively. These authors advocate instead ``the more conservative, model independent measure'' $\Delta_{\rm EW}$~\cite{Baer:2025hllhc}, which is generically expected to be the smallest of the three~\cite{Mustafayev:2014lqa}. A spectrum that is natural by $\Delta_{\rm EW}$ can therefore appear fine-tuned by the other two measures, and statements about naturalness below refer to $\Delta_{\rm EW}$.

Three conditions follow. First, the higgsino mass $|\mu|$ must lie near the weak scale, typically between 100 and 300 GeV. Second, $m_{H_u}^2$ must be driven to small negative values at the weak scale. Renormalization-group running with the large top Yukawa coupling drives it negative, which is the mechanism of radiative electroweak symmetry breaking~\cite{Ibanez:1982fr,Alvarez-Gaume:1983drc,Ellis:1983bp}. Because the breaking requires a large top Yukawa coupling, these analyses implied a heavy top quark well before its discovery. Ovrut and Raby found that it occurs for $m_t$ between 100 and 200 GeV~\cite{Ovrut:1983nb}, and that light color-octet and weak-triplet superfields lower this prediction by at least 20 GeV~\cite{Ovrut:1984gi}.

Third, the radiative corrections $\Sigma_u^u$ must be small. The largest of them come from the top squarks~\cite{Baer:2012cf},

\begin{equation}
\Sigma_u^u(\tilde t_{1,2})=\frac{3}{16\pi^2}\,F(m_{\tilde t_{1,2}}^2)\left[y_t^2-g_Z^2\mp\frac{y_t^2A_t^2-8g_Z^2\left(\frac14-\frac23x_W\right)\Delta_t}{m_{\tilde t_2}^2-m_{\tilde t_1}^2}\right],
\label{eq:sigmauu}
\end{equation}

where $y_t$ is the top Yukawa coupling of the MSSM and $F(m^2)=m^2\left[\ln(m^2/Q^2)-1\right]$ with $Q^2=m_{\tilde t_1}m_{\tilde t_2}$. The electroweak parameters are $g_Z^2=(g^2+g'^2)/8$, with $g$ and $g'$ the SU(2) and U(1) gauge couplings, and $x_W=\sin^2\theta_W$, with $\theta_W$ the weak mixing angle. Finally, $\Delta_t=(m_{\tilde t_L}^2-m_{\tilde t_R}^2)/2+M_Z^2\cos2\beta\left(\frac14-\frac23x_W\right)$, where $m_{\tilde t_L}^2$ and $m_{\tilde t_R}^2$ are the soft-breaking mass parameters of the left- and right-handed top squarks.

\subsection{Radiative natural supersymmetry}\label{sec:rns}

A large trilinear coupling $|A_t|$ raises $m_h$ toward its maximal-mixing value in Equation~(\ref{eq:mhsusy}) and also suppresses the top-squark terms in $\Sigma_u^u$. For the lighter top squark, the terms in the square brackets of Equation~(\ref{eq:sigmauu}) cancel as $|A_t|$ grows. For the heavier one, the factor $F$ is small when the two top squarks are well split, since $\ln(m_{\tilde t_2}^2/Q^2)=\ln(m_{\tilde t_2}/m_{\tilde t_1})$ is then close to 1~\cite{Baer:2012cf}. The mechanism that lifts the Higgs mass to 125 GeV also keeps the weak scale natural as measured by $\Delta_{\rm EW}$.

The cancellation at large $|A_t|$ underlies radiative natural supersymmetry (RNS), developed by Baer, Barger, and collaborators~\cite{Baer:2012up,Baer:2012cf}. RNS is realized within the MSSM with non-universal Higgs soft masses. A grand-unification-scale value of $m_{H_u}$ above the common scalar mass $m_0$ lets renormalization-group running, in which the top Yukawa coupling plays the main role, drive $m_{H_u}^2$ to a small negative value at the weak scale. Together with a large negative trilinear coupling, $A_0\approx-1.6\,m_0$, this yields $m_h\simeq125$ GeV and electroweak fine-tuning at the 3 to 10\% level (i.e., $\Delta_{\rm EW}\approx10$ to 30) with TeV-scale top squarks. First- and second-generation scalars may lie in the 10 to 30 TeV range, which helps with the supersymmetric flavor and CP problems~\cite{Baer:2012cf}.

The model predicts light higgsinos, with masses near $|\mu|$. Their small mass splittings make them hard to see at the LHC, but they would be produced directly at a linear $e^+e^-$ collider operating above their pair threshold~\cite{Baer:2012cf}. Natural solutions of the related non-universal Higgs model NUHM4 typically have the lighter top squark, $\tilde t_1$, between 1 and 2 TeV, within reach of the HL-LHC (Section~\ref{sec:susyfourtop})~\cite{Baer:2026pmssm,Baer:2025hllhc}. The thermal relic density of higgsino-like dark matter falls below the measured dark-matter density. If the strong CP problem is solved by the Peccei--Quinn mechanism, the dark matter can be a mixture of axions and higgsinos~\cite{Baer:2012cf}. In a string-landscape picture that favors large soft terms, subject to a weak scale of the observed size, the soft parameters are pulled toward the region of radiatively driven naturalness, where the Higgs mass is about 125 GeV~\cite{Baer:2016lpj,Baer:2017uvn}. The lighter top squark then lies between 1 and 2.5 TeV and decays comparably to $b\tilde\chi^+_1$ and to $t\tilde\chi^0_{1,2}$, so that the signal is spread over the $b\bar b$, $t\bar b$, and $t\bar t$ final states with missing transverse momentum. Baer, Barger, Dutta, Sengupta, and Zhang find an HL-LHC discovery reach of about 1.7 TeV and a 95\% CL exclusion reach of about 2 TeV in $m_{\tilde t_1}$, which cover most, but not all, of the natural landscape parameter space~\cite{Baer:2023tsq}.

In supersymmetric Higgs physics the top Yukawa coupling drives electroweak symmetry breaking, top and top-squark loops set the Higgs mass, and the top squarks largely control the fine-tuning. A precise top mass in a well-defined scheme, as expected from an $e^+e^-$ threshold scan, sharpens all three connections.

\section{Multi-Top Production}\label{sec:multitop}

Final states with three and four top quarks are among the rarest Standard Model processes accessible at the LHC. They probe the top sector in a regime complementary to the threshold. Four-top production is now observed, and three-top production is the target of a first dedicated search. Both channels respond to the size and phase of the top Yukawa coupling, to heavy states that couple mainly to top quarks, and to four-heavy-quark contact interactions.

\subsection{Four-top production}\label{sec:fourtop}

Four-top production proceeds mainly through $gg$ and $q\bar q$ initial states at order $\alpha_s^4$, with sizable electroweak and Higgs-exchange contributions. Table~\ref{tab:multitop} lists the predictions at NLO in QCD and the electroweak coupling~\cite{Frederix:2017wme} and with threshold resummation at next-to-leading-logarithmic accuracy including the one-loop constant terms (NLL$'$), which halves the scale uncertainty~\cite{vanBeekveld:2022hty}. ATLAS and CMS have observed four-top production at 6.1 and 5.6 standard deviations~\cite{ATLAS:2023ajo,CMS:2023ftu}. The measured cross sections lie above the central predictions, but not significantly so. The modeling of the main background, $\ttbar$ production with additional $b$ jets, is itself difficult. The measured $\ttbar b\bar b$ cross sections tend to lie above NLO predictions matched to parton showers, although mostly within the uncertainties~\cite{ATLAS:2018fwl,CMS:2023xjh}.

\subsection{Three-top production}\label{sec:threetop}

Three-top production requires a weak interaction, so the Standard Model rate is small. Barger, Keung, and Yencho found a Standard Model $tt\bar t$ plus $t\bar t\bar t$ cross section of about 2 fb at $\sqrt s=14$ TeV and showed that new physics could raise the rate by an order of magnitude or more~\cite{Barger:2010uw}. Gluino pair production and decay in the MSSM gives about 41 fb, and a leptophobic $Z'$ model gives about 28 fb. With efficient top tagging, the three-top signal could then provide evidence for new physics.

CMS has since performed the first dedicated search, using 138 fb$^{-1}$ in the single-lepton, same-sign dilepton, and multilepton channels~\cite{CMS:2025ttt}. It finds no significant excess and sets an observed upper limit of 25 fb at 95\% confidence level. The limit is an order of magnitude above the Standard Model rate, so the channel now constrains only the enhanced scenarios.

Beyond leading order, three-top and four-top production cannot be separated cleanly. Three-top production with a $W$ boson and an additional $b$ quark from radiation yields the same final state as four-top production followed by $t\to bW$. Durieux \emph{et al.} computed $tttW$ production at complete NLO~\cite{Durieux:2026ttt}. They find about 0.7 fb at 13 TeV for $\bar tt\bar tW^+$, i.e., about 1.4 fb summed over both charge states, compared with 12 to 15 fb for four-top production (Table~\ref{tab:multitop}). They argue that only a joint prediction should be compared with four-top measurements. The joint prediction exceeds the purely on-shell four-top rate by more than 10\%. A consistent treatment of this overlap is a prerequisite for interpreting any future three-top signal.

\begin{table}[H]
\centering
\caption{Multi-top production at the LHC. Cross sections are for $\sqrt s=13$ TeV unless noted. The value of Ref.~\cite{Durieux:2026ttt} is for the $tttW$ subprocess only, summed over both charge states. The NLO+NLL$'$ value is from the revised version of Ref.~\cite{vanBeekveld:2022hty} (arXiv v2, 2025), which supersedes the published value of 13.37 fb. For CMS four-top production the statistical and systematic uncertainties are combined in quadrature.}
\label{tab:multitop}
\begin{tabular}{lll}
\toprule
Process & Standard Model prediction & Measurement \\
\midrule
$t\bar t t\bar t$ & $12.0^{+2.2}_{-2.5}$ fb (NLO QCD+EW)~\cite{Frederix:2017wme} & $22.5^{+6.6}_{-5.5}$ fb, $6.1\sigma$ (ATLAS)~\cite{ATLAS:2023ajo} \\
 & $14.65^{+1.57}_{-2.75}$ fb (NLO+NLL$'$)~\cite{vanBeekveld:2022hty} & $17.7^{+4.4}_{-4.0}$ fb, $5.6\sigma$ (CMS)~\cite{CMS:2023ftu} \\
$tt\bar t+t\bar t\bar t$ & $\approx 2$ fb (14 TeV, LO)~\cite{Barger:2010uw} & $<25$ fb, 95\% CL (CMS)~\cite{CMS:2025ttt} \\
$tttW$ & $\approx 1.4$ fb (complete NLO)~\cite{Durieux:2026ttt} & \\
\bottomrule
\end{tabular}
\end{table}

\subsection{The top Yukawa coupling and the Higgs width}\label{sec:multitopyukawa}

Four-top production bounds the top Yukawa coupling and, combined with Higgs-boson measurements, the Higgs width, under assumptions stated below. The four-top amplitudes include diagrams in which an off-shell Higgs boson is exchanged between top-quark lines. These diagrams contribute terms proportional to $y_t^2$ through interference with the QCD and electroweak gauge amplitudes, and terms proportional to $y_t^4$ from the Higgs-exchange amplitude squared. They do not depend on the Higgs width. In terms of the CP-even and CP-odd modifiers $\kappa_t=g_{Htt}\cos\xi$ and $\tilde\kappa_t=g_{Htt}\sin\xi$ of Equation~(\ref{eq:cpyukawa}), CMS gives the leading-order parameterization at 13 TeV~\cite{CMS:2026mt}

\begin{equation}
\sigma_{\rm LO}(t\bar tt\bar t)=\left(6.4-0.8\kappa_t^2+1.9\tilde\kappa_t^2+0.74\kappa_t^4+1.8\kappa_t^2\tilde\kappa_t^2+1.2\tilde\kappa_t^4\right)~{\rm fb},
\end{equation}

shown in Figure~\ref{fig:kappa}. The negative $\kappa_t^2$ term reflects destructive interference between Higgs exchange and the QCD and electroweak gauge amplitudes~\cite{Cao:2016wib,Cao:2019ygh}. Only even powers of the modifiers appear.

The on-shell rate for $t\bar tH$ followed by a given Higgs decay, by contrast, is proportional to $y_t^2$ times a branching ratio that depends on the Higgs width $\Gamma_H$. Combining the two can constrain both $y_t$ and $\Gamma_H$ without assumptions about unobserved Higgs decays, as pointed out by Cao, Chen, and Liu~\cite{Cao:2016wib}. ATLAS has applied this method to its Higgs-boson and four-top measurements and bounds $\Gamma_H<450$ MeV at 95\% confidence level, assuming the same top Yukawa coupling in on-shell and off-shell Higgs processes~\cite{ATLAS:2024hwidth}. The bounds on $\kappa_t$ below instead treat the $t\bar tH$ background with Standard Model Higgs decays.

ATLAS used its four-top measurement to bound the top Yukawa coupling below 1.9 times its Standard Model value at 95\% confidence level~\cite{ATLAS:2023ajo}. A combined CMS analysis of four- and three-top events measures $\sigma(t\bar tt\bar t)=18.9^{+4.4}_{-4.0}$ fb. It finds $0.7<|\kappa_t|<1.4$ at 95\% confidence level for $\tilde\kappa_t=0$ (expected $0.6<|\kappa_t|<1.3$), and a 30\% uncertainty on the $t\bar tH$ rate, covering the experimental uncertainty on $\Gamma_H$, allows values of $|\kappa_t|$ up to 13\% higher~\cite{CMS:2026mt}. With the $t\bar tH$ normalization left free, the ATLAS bound becomes 2.3~\cite{ATLAS:2023ajo}. Four-top production alone was projected to exclude a purely CP-odd coupling at the 13 TeV LHC with about 680 fb$^{-1}$~\cite{Cao:2019ygh}.

\begin{figure}[H]
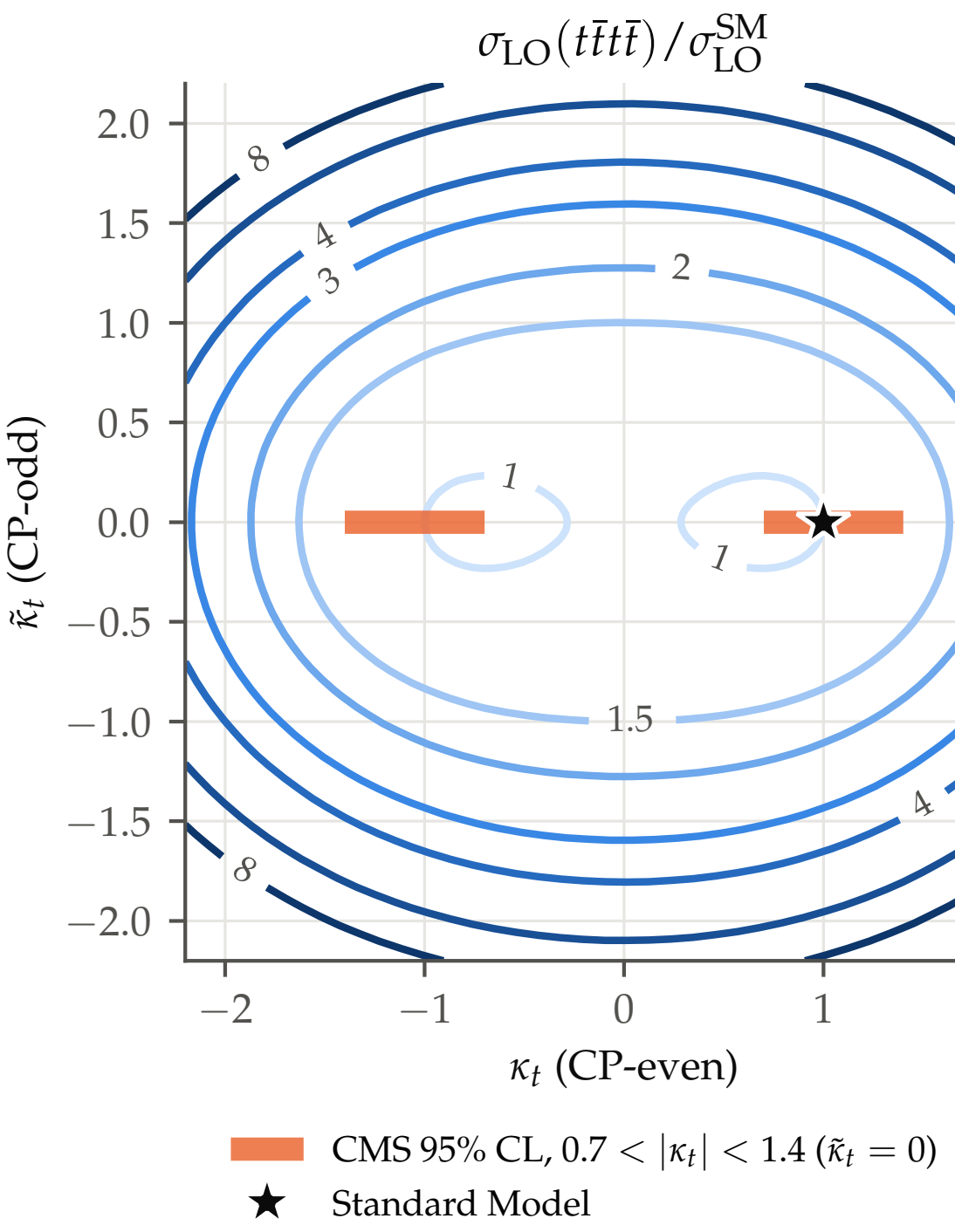

\centering
% ==== embedded figure fig_kappa: generated by make_figures.py (matplotlib PGF backend); regenerate rather than edit ====
\begingroup%
\makeatletter%
%
\makeatother%
\endgroup%
% ==== end of embedded figure fig_kappa ====
\caption{Contours of the leading-order four-top cross section relative to its Standard Model value in the plane of the CP-even and CP-odd top Yukawa modifiers, from the CMS parameterization quoted in the text~\cite{CMS:2026mt}. The orange bands show the CMS 95\% CL interval $0.7<|\kappa_t|<1.4$ for $\tilde\kappa_t=0$, and the star marks the Standard Model. The CMS interval also uses the three-top channels and higher-order normalizations, so it cannot be read off these contours alone. Only even powers of the modifiers enter, and four-top production does not determine the sign of $\kappa_t$.}
\label{fig:kappa}
\end{figure}

Unlike the four-top rate, the three-top cross sections contain terms linear in $\kappa_t$, which CMS traces to diagrams with a single $Ht\bar t$ vertex, and the $tttq$ cross section is strongly enhanced for negative $\kappa_t$~\cite{CMS:2026mt}. A term linear in $\kappa_t$ requires interference between an amplitude with one $Ht\bar t$ vertex and an amplitude that does not involve the top Yukawa coupling. The same structure governs $tHj$ production. In the unitary gauge the $Ht\bar t$ amplitude interferes destructively with the $HWW$ amplitude in the Standard Model, and a negative $\kappa_t$ makes the interference constructive and enhances the rate (Section~\ref{sec:interference})~\cite{Barger:2018tqn,Barger:2019ccj}. Three-top production thus joins $tHj$ as a probe of the sign of $\kappa_t$, although at present rates its statistical power is limited.

\subsection{New heavy states and contact interactions}\label{sec:heavystates}

A heavy scalar, pseudoscalar, or vector boson that couples mainly to top quarks, in either a color-singlet or color-octet representation, can be produced with a $t\bar t$ pair and decay to $t\bar t$, giving four tops in the final state. CMS has interpreted its Run~2 four- and three-top data in terms of such resonances. It constrains such states with masses between 400 GeV and 1.6 TeV, depending on spin and color. It also constrains SMEFT four-heavy-quark operators and the top-Higgs operator~\cite{CMS:2026mt}.

Four-top production complements the threshold analysis of Section~\ref{sec:heavyA}. Suppose that a pseudoscalar with mass above $2m_t$ couples mainly to top quarks and distorts the $\mtt$ spectrum through interference in $gg\to A\to t\bar t$. Such a state would also be produced in $t\bar t A$ associated production, where it appears as a peak in a $t\bar t$ pair recoiling against a second pair. Four-heavy-quark operators enter four-top production at tree level and Higgs and electroweak precision observables at loop level. A global fit that combines these data with $\ttbar$ production lifts flat directions of the four-top data and reduces the allowed parameter space~\cite{DiNoi:2025fhq}.

\subsection{Supersymmetric four-top signals}\label{sec:susyfourtop}

In supersymmetric models in which the top squarks are the lightest squarks, as in RNS (Section~\ref{sec:rns}), the gluino $\tilde g$ decays mainly to a top quark and an on- or off-shell top squark~\cite{Baer:2025hllhc}. The primary final states of heavy-gluino decay~\cite{Baer:1986au,Baer:1991xs,Baer:1998bj} are then $\tilde g\to t\bar t\tilde\chi^0_1$ and $\tilde g\to t\bar b\tilde\chi^-_1+{\rm c.c.}$, with $\tilde\chi^0_1$ the lightest neutralino and $\tilde\chi^\pm_1$ the lighter chargino. Gluino pair production can therefore provide a new-physics source of triple- and quadruple-top events, accompanied by large missing transverse momentum from the two neutralinos. The final state has up to four $b$ jets, and the missing momentum separates it from Standard Model four-top production. Gluino decays were also among the enhanced three-top scenarios of Ref.~\cite{Barger:2010uw}.

The $t\bar t$ pair from a single gluino decay cannot form the $\etat$. The gluino is a color octet and the neutralino a singlet, so the $t\bar t$ pair is in a color-octet state and feels a repulsive potential. The relevant bound-state physics is instead the enhancement of gluino-pair production near threshold from color-singlet $\tilde g\tilde g$ (gluinonium) states~\cite{Hagiwara:2009hq}.

With the full Run~2 data, ATLAS excludes gluino masses below 2.44 TeV for a massless neutralino in this decay mode, using final states with three or more $b$ jets~\cite{ATLAS:2022gtt}. CMS excludes masses up to about 2.15 TeV for neutralino masses up to 700 GeV in single-lepton final states~\cite{CMS:2019gtt}. The sensitivity falls when the spectrum is compressed, i.e., when $m_{\tilde g}-m_{\tilde\chi^0_1}$ approaches $2m_t$, because the top quarks are soft and the missing momentum is small. Searches in this region could use events with hard initial-state radiation and improved reconstruction of soft top-quark decays.

Top squarks are the central states in this picture. For top squarks from the landscape, the HL-LHC reach in $m_{\tilde t_1}$ is about 1.7 TeV for discovery and about 2 TeV for exclusion (Section~\ref{sec:rns})~\cite{Baer:2023tsq}, and the study of natural NUHM4 solutions by Baer, Barger, and collaborators~\cite{Baer:2026pmssm} finds that combining top-squark searches with higgsino and wino pair-production searches could explore much of the natural parameter space. The HL-LHC prospects for supersymmetry, including gluino searches in final states with multiple $b$-tagged jets, are reviewed in Ref.~\cite{Baer:2025hllhc}, and the HL-LHC reach for landscape supersymmetry is compared across signatures in the plane of the higgsino and gaugino mass parameters in Ref.~\cite{Baer:2026ytu}. The Standard Model four-top measurement and the supersymmetric search are complementary. The former fixes the irreducible background and its modeling, and the latter probes the tail at high mass and high missing momentum.

\subsection{Outlook}\label{sec:multitopoutlook}

With roughly twenty times the Run~2 luminosity, the HL-LHC will improve the four-top measurement, whose rate is sensitive to $y_t^4$, and extend three-top searches toward the enhanced scenarios. For gluino pair production with decays to quarks and a massless bino-like neutralino, it is projected to probe gluino masses up to about 3.2 TeV~\cite{CidVidal:2018eel}. In RNS, where the gluino decays through the top squark to third-generation quarks and soft higgsino decay products, the projected reach is 2.4 TeV with 300 fb$^{-1}$ and 2.8 TeV with 3 ab$^{-1}$, and the rate of the selected gluino events determines $m_{\tilde g}$ to 2 to 5\%~\cite{Baer:2016wkz}.

\section{Future Directions}\label{sec:future}

\subsection{Hadron colliders}\label{sec:hadroncolliders}

Run~3 ended in May 2026 with more than twice the Run~2 integrated luminosity at 13.6 TeV, so that ATLAS and CMS have each recorded about 0.5 ab$^{-1}$ over the three LHC runs~\cite{CERN:2026run3}. To our knowledge no threshold analysis of the Run~3 data has yet been published, although part of them has entered other top-quark analyses (e.g., 164 fb$^{-1}$ in Ref.~\cite{ATLAS:2026cpgg}). With these data and the 3 ab$^{-1}$ expected at the HL-LHC~\cite{CidVidal:2018eel}, the statistical uncertainty on the $\etat$ rate will become small compared with the modeling uncertainties of Section~\ref{sec:model}. Progress on the cross section will then depend mainly on theory.

The Run~3 and HL-LHC data will not resolve the internal structure of the threshold, because the $\mtt$ and velocity resolutions are too coarse. They will, however, allow measurements of the spin and parity content of the excess and measurements in several channels with matched signal definitions~\cite{CMS:2026lj,Fuks:2025single}. At high mass the same data will sharpen four-top measurements and extend searches for four tops with missing transverse momentum from gluino decays (Section~\ref{sec:susyfourtop}). The vector state $\psi_t$ is difficult to observe at the LHC because of the selection rule of Section~\ref{sec:channels} that forbids its production from two gluons at leading order, and because of the gluon-dominated environment~\cite{Bai:2025toponia}.

A hadron collider at 85 to 100 TeV, the second stage of the Future Circular Collider (FCC) program~\cite{FCC:2025uan,FCC:2025lpp}, would extend the LHC top-quark program to higher energy. Using the ratio of $\ttbar H$ to $\ttbar Z$ production, with the $Zt\bar t$ couplings fixed at FCC-ee, the $e^+e^-$ first stage, it could measure $y_t$ at the percent level~\cite{Mangano:2015aow,FCC:2025lpp}. Its direct reach for new heavy states, including those coupled mainly to top quarks, would extend to masses of tens of TeV~\cite{FCC:2025lpp}.

\subsection{Lepton colliders}\label{sec:leptoncolliders}

A detailed study for FCC-ee projects experimental precisions of 6.8 MeV on $m_t$ and 11.5 MeV on $\Gamma_t$ from the threshold scan, with $m_t$ defined in the potential-subtracted scheme~\cite{Defranchis:2025tth}. The theory uncertainty from missing higher orders is about 35 MeV on $m_t$ and 25 MeV on $\Gamma_t$. At an $e^+e^-$ collider the threshold is produced through the vector current, the $\psi_t$ channel is accessible, and the initial state is known. The $e^+e^-\to\ttbar$ threshold cross section is known to N$^3$LO in QCD~\cite{Beneke:2015kwa}.

The threshold scan is also sensitive to $y_t$ and $\alpha_s$. These produce similar shape effects and must be disentangled. With $y_t$ left free, the scan determines it to about 6\%. A dedicated run above threshold, at 365 GeV, would reach an experimental uncertainty of about 2\%. A prediction of the cross section accurate to 0.1\% would add a theoretical uncertainty of similar size, for a total of about 3\%~\cite{Defranchis:2025tth}.

A toponium signal in the ratio of $b\bar b$ to inclusive hadronic cross sections at the Circular Electron Positron Collider (CEPC) has also been proposed~\cite{Fu:2024bki}. Because the $\etat$ and $\psi_t$ are nearly degenerate, a measurement of the position of the $\etat$ enhancement at the HL-LHC would help choose the center-of-mass energies of such a scan. The $\psi_t$ should be discoverable at an $e^+e^-$ collider both through its constituent decays to $W^+W^-b\bar b$ and through the interference of its two-body annihilation decays with the continuum~\cite{Bai:2025toponia}.

A muon collider offers two further options. Run at the $\ttbar$ threshold, it would have a naturally narrow beam energy spread, which suits a scan of the threshold lineshape~\cite{Berger:1995se}. An early study by Barger, Berger, Gunion, and Han estimated a top-mass precision of about 200 MeV with 10 fb$^{-1}$ and 70 MeV with 100 fb$^{-1}$, provided the systematic uncertainties are understood~\cite{Barger:1997yk}. At multi-TeV energies a muon collider would instead probe the top quark through vector-boson fusion, including the CP phase of the top Yukawa coupling (Section~\ref{sec:muoncollider})~\cite{Cassidy:2023lwd}, and through $\ttbar$ and $\ttbar H$ production, which is sensitive to the electroweak dipole moments of the top quark (Section~\ref{sec:cpfuture})~\cite{Han:2024tdm}.

\subsection{Toponium spectroscopy}\label{sec:spectroscopy}

In the 1980s toponium was expected to have narrow levels, radiative transitions, and annihilation decays that could serve as a source of Higgs bosons~\cite{Kuhn:1987ty,Barger:1987kk}, as for heavy quarkonium in general~\cite{Barger:1986nn}. It was also recognized early that toponium levels would add little to the determination of the QCD scale parameter from heavy-quarkonium spectra, because the fitted scale is correlated with the quark masses~\cite{Hagiwara:1983qw}. For the actual top mass every level acquires the width $2\Gamma_t\simeq2.7$ GeV from the weak decay of either constituent. This exceeds the 1S--2S separation of $3|E_1|/4\simeq1$--$1.3$ GeV in the Coulomb approximation of Equation~(\ref{eq:hydrogen}). The radial excitations merge into the threshold continuum (Figure~\ref{fig:green}b). Whether distinct states exist is a question about poles of the $t\bar t$ scattering amplitude in the complex energy plane~\cite{Tang:2026cep}.

Annihilation decays are strongly suppressed relative to the weak decay of a constituent top quark. At leading order $\Gamma(\eta_t\to gg)=8\alpha_s^2|R_{1S}(0)|^2/(3M^2)$. Evaluated with the Coulomb wave function at the origin, $|R_{1S}(0)|^2=4/r_B^3$ with $r_B$ of Equation~(\ref{eq:hydrogen}), and with $\alpha_s(m_t)$, this width is a few MeV, so annihilation branching ratios are of order $10^{-3}$ or smaller. Reference~\cite{Wang:2024toponia} computes such annihilation widths and finds the spin-singlet and spin-triplet levels essentially degenerate at this mass. Toponium spectroscopy means determining the spin, parity, and color composition of the threshold enhancement, and comparing rates and lineshapes in different production channels, since individual lines cannot be resolved. Potential-model and Bethe--Salpeter calculations of the spectrum and wave functions~\cite{Wang:2024toponia,Bai:2025toponia} and the third-order potential corrections to the S-wave Green's function~\cite{Beneke:2024threshold} supply the theoretical input.

\begin{table}[H]
\centering
\caption{Low-lying color-singlet toponium states and their production at leading order in different initial states. ``Suppressed'' denotes P-wave production, which vanishes at threshold. For $\chi_{t1}$ in $gg$ and $\gamma\gamma$ the Landau--Yang theorem forbids production, and in $e^+e^-$ it proceeds through the axial-vector coupling of the $Z$ boson.}
\label{tab:spectroscopy}
\begin{tabular}{lcccc}
\toprule
State & $J^{PC}$ & $gg$ (LHC) & $e^+e^-$ & $\gamma\gamma$ \\
\midrule
$\eta_t$ (${}^1S_0$) & $0^{-+}$ & dominant & no & yes \\
$\psi_t$ (${}^3S_1$) & $1^{--}$ & no & dominant & no \\
$\chi_{t0}$ (${}^3P_0$) & $0^{++}$ & suppressed & no & suppressed \\
$h_t$ (${}^1P_1$) & $1^{+-}$ & no & no & no \\
$\chi_{t1}$ (${}^3P_1$) & $1^{++}$ & no & suppressed ($Z$ axial) & no \\
$\chi_{t2}$ (${}^3P_2$) & $2^{++}$ & suppressed & no & suppressed \\
\bottomrule
\end{tabular}
\end{table}

Different initial states select different members of the spectrum (Figure~\ref{fig:levels} and Table~\ref{tab:spectroscopy}). At the LHC gluon fusion produces the $\etat$, together with P-wave $\chi_{t0}$ and $\chi_{t2}$ contributions that vanish at threshold. Scalar or vector admixtures in the excess can be tested with the Run~3 and HL-LHC data, using the spin-triplet observable $c_{\rm han}$~\cite{CMS:2025kzt,CMS:2026lj} and the decay-plane correlation of Equation~(\ref{eq:PhiH})~\cite{Barger:2011pu}. An annihilation decay, $\etat\to W^+W^-$ with dilepton final states free of $b$ jets, was proposed in Ref.~\cite{Yang:2024hnd} and studied as a distinct signature at the HL-LHC~\cite{Yang:2026etaww}. The P-wave states remain difficult to observe at both the HL-LHC and future lepton colliders~\cite{Bai:2025toponia}.

At an $e^+e^-$ collider the threshold scan maps the smeared $\psi_t$ lineshape, and the axial-vector $Z$ coupling adds a P-wave $\chi_{t1}$ component~\cite{Bai:2025toponia,Defranchis:2025tth}. The $\etat$ at the LHC and the $\psi_t$ at a lepton collider are governed by the same color-singlet S-wave Green's function. Comparing the two channels tests the threshold description.

Photon fusion gives independent access to the $\etat$ (Table~\ref{tab:spectroscopy}). The photons carry no color, so the $\ttbar$ pair is produced only in a color-singlet state. In ultraperipheral collisions the protons or nuclei pass at large impact parameter, interact through their photon fields, and remain intact. The exclusive cross section for $\gamma\gamma\to\etat$ in such collisions has been computed for proton--proton, proton--lead, and lead--lead beams at LHC and FCC energies by Francener, Gon\c{c}alves, and Martins~\cite{Francener:2025tor} and by d'Enterria and Kang~\cite{dEnterria:2025ecx}. It is proportional to the two-photon width of the $\etat$, taken as 7.6 or 57 keV in Ref.~\cite{Francener:2025tor} and as 27 keV, summed over the S-wave levels, in Ref.~\cite{dEnterria:2025ecx}. For proton--proton collisions at 14 TeV the results are 1.5 to 13 ab~\cite{Francener:2025tor} and $6.4\pm2.0$ ab~\cite{dEnterria:2025ecx}, which agree once the widths are taken into account. Lead--lead collisions at the LHC have larger cross sections, about 1 fb in Ref.~\cite{dEnterria:2025ecx} and 5 to 41 fb in Ref.~\cite{Francener:2025tor}. Either value gives much less than one event for the 10 nb$^{-1}$ assumed in Ref.~\cite{dEnterria:2025ecx}.

The conclusion of d'Enterria and Kang is that the measurement can be attempted only in proton--proton collisions. About 40 $\etat$ would be produced in 6 ab$^{-1}$ at the HL-LHC (twice the 3 ab$^{-1}$ expected for each experiment) and about 1300 in 30 ab$^{-1}$ at a 100 TeV collider~\cite{dEnterria:2025ecx}. The proposed signature is a $\ttbar$ pair with small total transverse momentum in coincidence with two intact protons detected in forward spectrometers. These yields are numbers of produced events, before the top-quark decay branching fractions, acceptance, and efficiency are applied. The $\gamma\gamma\to\ttbar$ continuum, about 300 ab at the LHC when integrated over $\mtt$, is about 50 times the $\etat$ rate~\cite{dEnterria:2025ecx}. The CMS and TOTEM search for exclusive $\ttbar$ production with tagged protons sets a limit of 0.59 pb at 95\% CL, three orders of magnitude above this continuum~\cite{CMS:2023naq}.

Other photon-induced channels have much smaller rates. Li, Liang, and Yang compute $\gamma\gamma\to\psi_t\gamma$, in which the vector state is produced together with a photon, and find about 5 ab in lead--lead collisions at $\sqrt{s_{\rm NN}}=39.4$ TeV~\cite{Li:2026upc}. The cross section for $\gamma\gamma\to\etat$ at that energy is of order 10 to 100 pb~\cite{Francener:2025tor,dEnterria:2025ecx}. For $\mu^+\mu^-$ annihilation to toponium and a photon at 3 and 10 TeV, the same authors find cross sections well below 1 ab~\cite{Li:2026upc}.

\subsection{CP violation in the top sector}\label{sec:cpfuture}

Because the top Yukawa coupling is of order one, a modest CP-violating phase $\xi$ in Equation~(\ref{eq:cpyukawa}) would be a sizable new source of CP violation, of the kind required to generate the baryon asymmetry in electroweak baryogenesis~\cite{Morrissey:2012db}. Present LHC data allow $|\xi|$ up to about $38^\circ$~\cite{ATLAS:2026cpgg}, so the phase is far from determined.

The direct program combines three classes of observables. Angular and kinematic analyses of $t\bar tH$ production will use the full HL-LHC data set. Interference-based probes are sensitive to the relative phase of the $Ht\bar t$ and gauge couplings. These include $tHj$ production at the LHC~\cite{Barger:2018tqn,Barger:2019ccj}, the three-top channels (Section~\ref{sec:multitopyukawa})~\cite{CMS:2026mt}, and vector-boson fusion at a multi-TeV muon collider (Section~\ref{sec:muoncollider})~\cite{Cassidy:2023lwd}. Measurements at an $e^+e^-$ collider complete the set~\cite{Hagiwara:2017ban}. The SMEFT ties these measurements together, because the dimension-six operator that generates $\xi$ also generates $t\bar tHH$ interactions whose high-energy behavior can be tested~\cite{Barger:2023wbg}.

Electric dipole moments (EDMs) give indirect constraints. The electron EDM bound $|d_e|<4.1\times10^{-30}\,e\,$cm at 90\% confidence level~\cite{Roussy:2022cmp} constrains a CP-odd top Yukawa coupling through two-loop contributions. The resulting limit, however, assumes the Standard Model Higgs coupling to the electron and disappears without it. The neutron and mercury EDM limits likewise assume Standard Model couplings to light quarks and allow $\tilde\kappa_t$ of order one without them~\cite{Brod:2013cka}. Direct collider measurements avoid this assumption.

CP violation can also enter through the chromoelectric and electroweak dipole moments of the top quark. At the LHC these are probed through T-odd triple-product asymmetries in $t\bar t$ events, where CMS finds results consistent with the Standard Model~\cite{CMS:2022cpvtt}, and through CP-odd observables in top-quark production with a $Z$ boson~\cite{CMS:2025tzcp}. Lepton colliders, both $e^+e^-$ machines running at and above the $t\bar t$ threshold and a high-energy muon collider through $t\bar t$ and $t\bar tH$ production, offer cleaner access to the electroweak dipoles~\cite{Han:2024tdm}. Finally, the threshold region itself is a natural place to look for CP-odd couplings of new states, through the phase and the $\sin\Phi$ term of the decay-plane modulation of Equation~(\ref{eq:PhiH})~\cite{Barger:2011pu}.

\subsection{Theory}\label{sec:theoryfuture}

Four theoretical developments would improve the comparison of the threshold data with theory (Section~\ref{sec:missing}).
\begin{itemize}
\item A matched combination of NRQCD resummation with NNLO perturbative QCD, including off-shell decays.
\item Inclusion of P-wave, color-octet, and quark-initiated contributions with full spin correlations.
\item Propagation of the $m_t$ and scheme dependence of the Green's function.
\item A consistent treatment of new-physics amplitudes near threshold.
\end{itemize}

\section{Conclusions}\label{sec:concl}

The LHC has observed the threshold enhancement of top-pair production that NRQCD predicts. The enhancement carries the angular signature of a spin-singlet state and is consistent with the color-singlet $\etat$. It is established by two experiments and in two decay channels, each above five standard deviations.

The measured cross sections depend on the signal model used to extract them, and with the same model they still differ between analyses by factors of about 2.6 (simplified $\etat$ model) and 3.3 (Green's-function model). These two effects are the main obstacle to a precise comparison with theory. The Green's-function description developed by Fuks, Hagiwara, Ma, and Zheng is the nominal model of ATLAS and an alternative model in the CMS lepton+jets analysis. The next step is to extend it to higher orders and other partial waves and to match it to perturbative baselines. Generators that combine the resummed threshold corrections with NLO perturbative QCD, including off-shell effects, are already available (Section~\ref{sec:missing}). With the Run~3 and HL-LHC data, the precision will be set by this theory work.

The threshold enhancement, modeled as toponium or through higher orders in $\alpha_s/v$, must now be included in analyses of the threshold region. These include measurements of spin entanglement, extractions of the top mass, global effective-field-theory fits, and searches for heavy pseudoscalars coupled to the top quark. In those searches the choice between a pure Standard Model effect and an additional state that couples mainly to top quarks remains open.

The threshold also bears on the meaning of the top mass. Direct measurements determine the mass parameter of the event generators, while cross sections determine the pole and $\overline{\rm MS}$ masses. CMS estimated that Coulomb and soft-gluon resummation, which its extraction neglected, would raise the pole mass it extracted near threshold by about 1 GeV. The present pole-mass values range from about one standard deviation below to about 2.6 standard deviations above the Standard Model bound for absolute stability of the electroweak vacuum. The value below the bound is the CMS extraction, which the estimated resummation shift would move to the bound. Whether the Standard Model vacuum is stable or metastable therefore depends on how the mass is extracted and on the relation between the measured and pole masses.

Away from threshold, Standard Model predictions for the top quark continue to agree with precise measurements. The Tevatron forward-backward asymmetry, once read as a sign of new physics, agrees with the NNLO prediction, and the LHC charge asymmetry is seen at 4.7 standard deviations. Top-quark loops fix the side $R_t$ of the unitarity triangle, and the sides alone give $\gamma\simeq65^\circ$, in agreement with direct determinations from tree-level decays. In supersymmetry the top Yukawa coupling lifts the Higgs mass to 125 GeV, and radiative natural supersymmetry reconciles the required top squarks with low fine-tuning as measured by $\Delta_{\rm EW}$.

At higher mass, four-top production is established and three-top production is under direct search. Together they bound the top Yukawa coupling, are sensitive to its sign, and test the same heavy states that could appear at threshold. Combined with Higgs-boson measurements, four-top production also bounds the Higgs width. Boosted top quarks extend these tests to the highest masses, and top quarks produced in nuclear collisions could probe the time structure of the quark--gluon plasma once higher heavy-ion luminosities are available.

Looking further ahead, an $e^+e^-$ threshold scan would turn the same physics into determinations of $m_t$, in a short-distance scheme, and of $\Gamma_t$ to a few tens of MeV. A dedicated run above threshold could determine $y_t$ to about 3\% if the cross section is predicted to 0.1\%. A muon collider running at the threshold would offer a narrow beam energy spread for the same scan. An 85 to 100 TeV hadron collider would measure $y_t$ at the percent level. Interference-based probes at the LHC and at a muon collider could measure the CP phase of the top Yukawa coupling. The heaviest quark, long regarded as too short-lived to form an observable bound state, now gives access to bound-state QCD at short distances, although the present LHC data do not yet separate the resummed threshold dynamics from fixed-order QCD (Section~\ref{sec:implications}).

\vspace{6pt}

\authorcontributions{All aspects of this work, H.B. and V.B. Both authors contributed equally to this work and have read and agreed to the published version of the manuscript.}

\funding{This research received no external funding.}

\institutionalreview{Not applicable.}

\informedconsent{Not applicable.}

\dataavailability{No new data were created or analyzed in this study. The experimental results discussed are publicly available in the cited publications and in the Review of Particle Physics of the Particle Data Group~\cite{ParticleDataGroup:2026mpi}.}

\acknowledgments{We thank the guest editors, Ya-Juan Zheng and Ian Lewis, for the invitation to contribute to this Special Issue. V.B. thanks Kaoru Hagiwara and Ya-Juan Zheng for collaboration on the top Yukawa coupling. We thank Herbi Dreiner for comments on tests of locality and entanglement at colliders, Stefano Pozzorini for comments on the definition of the top-quark mass in event generators, and David d'Enterria for comments on photon-fusion production of toponium. During the preparation of this work the authors used Anthropic's Claude (Fable 5.1) in literature searches and editing language. When using this tool/service, the authors reviewed and edited the output and take full responsibility for the content of the published article.}

\conflictsofinterest{The authors declare no conflicts of interest.}

\abbreviations{Abbreviations}{
The following abbreviations are used in this manuscript:
\\

\noindent
\begin{tabular}{@{}ll}
CEPC & Circular Electron Positron Collider\\
CHSH & Clauser--Horne--Shimony--Holt\\
CKM & Cabibbo--Kobayashi--Maskawa\\
CL & confidence level\\
CP & charge conjugation and parity\\
CPT & charge conjugation, parity, and time reversal\\
EDM & electric dipole moment\\
FCC & Future Circular Collider (FCC-ee, its $e^+e^-$ stage)\\
GFRW & Green's-function reweighting\\
GIM & Glashow--Iliopoulos--Maiani\\
HL-LHC & High-Luminosity LHC\\
LEP & Large Electron--Positron collider\\
LHC & Large Hadron Collider\\
LO & leading order\\
MC & Monte Carlo\\
$\overline{\rm MS}$ & modified minimal subtraction\\
MSSM & minimal supersymmetric standard model\\
N$^3$LO & next-to-next-to-next-to-leading order\\
NLL & next-to-leading logarithmic\\
NLO & next-to-leading order\\
NNLO & next-to-next-to-leading order\\
NRQCD & nonrelativistic QCD\\
NUHM4 & non-universal Higgs mass model (version 4)\\
PDF & parton distribution function\\
PDG & Particle Data Group\\
PS & parton shower\\
QCD & quantum chromodynamics\\
RNS & radiative natural supersymmetry\\
SLC & SLAC Linear Collider\\
SMEFT & Standard Model effective field theory
\end{tabular}
}

\begin{adjustwidth}{-\extralength}{0cm}

\reftitle{References}

\PublishersNote{}
\end{adjustwidth}
\end{document}